\documentclass[11pt]{article}
\usepackage[a4paper, total={5in, 9in}]{geometry}
\usepackage{siunitx}
\usepackage{bm}
\usepackage{amsmath}
\usepackage{physics}
\usepackage{graphicx}
\usepackage{authblk}
\usepackage{natbib}

\newcommand\blfootnote[1]{%
  \begingroup
  \renewcommand\thefootnote{}\footnote{#1}%
  \addtocounter{footnote}{-1}%
  \endgroup
}

\setcitestyle{open={(},close={)}}

\title{Modelling the three-dimensional, diagnostic anisotropy field of an ice rise}
\author[1, 2, *]{A. Clara J. Henry}
\author[3]{Carlos Martín}
\author[2]{Reinhard Drews}
\affil[1]{Max Planck Institute for Meteorology, Hamburg, Germany}
\affil[2]{Department of Geosciences, University of Tübingen, Tübingen, Germany}
\affil[3]{British Antarctic Survey, Natural Environmental Research Council, Cambridge, UK}
\affil[*]{Now at: Department of Mathematics, Stockholm University, Sweden}
\date{}

\begin{document}

\maketitle

\section*{Abstract}

Polar ice develops anisotropic crystal orientation fabrics under deformation, yet ice is most often modelled as an isotropic fluid. We present three-dimensional simulations of the crystal orientation fabric of Derwael Ice Rise including the surrounding ice shelf using a crystal orientation tensor evolution equation corresponding to a fixed velocity field. We use a semi-Lagrangian numerical method that constrains the degree of crystal orientation evolution to solve the equations in complex flow areas. We perform four simulations based on previous studies, altering the rate of evolution of the crystal anisotropy and its dependence on a combination of the strain rate and deviatoric stress tensors. We provide a framework for comparison with radar observations of the anisotropy field, outlining areas where the assumption of one vertical eigenvector may not hold and provide resulting errors in measured eigenvalues. We recognise the areas of high horizontal divergence at the ends of the flow divide as important areas to make comparisons with observations. Here, poorly constrained model parameters result in the largest difference in fabric type. These results are important in the planning of future campaigns for gathering data to constrain model parameters and as a link between observations and computationally-efficient, simplified models of anisotropy. \blfootnote{Correspondence: clara.henry@math.su.se}

\section{Introduction}

Ice varies from isotropic to anisotropic, with crystals developing a preferred orientation fabric due to ice flow dynamics. Ice crystals tend to align with the direction of an applied force and shearing of ice is enhanced along basal planes perpendicular to the applied force \citep{Alley1988}. The orientation of the basal planes is described by a vector referred to as the $c$-axis. The snow grain orientation is initially isotropic and anisotropic ice fabric develops at a rate dependent on a variety of physical factors including the temperature, pressure and strain rate of the ice. These dependencies allow the investigation of both the steady-state and transient behaviour of ice through the analysis of observational anisotropy data and ice flow models.

Ice cores have long been used to quantify the the direction of ice crystals and the degree of anisotropy with depth \citep{Durand2009,Montagnat2014, Weikusat2017}. However, the lack of horizontal spatial information hinders the study of the relationship between ice flow and anisotropy. Apart from ice cores sites, anisotropic information is sparse and relies on geophysical methods such as seismics, both active \citep{Diez2014, Brisbourne2019} and passive \citep{Smith2017, Kufner2023} or radar \citep{Fujita2006, Matsuoka2012, Drews2012}. The main limitation is that, generally, seismic methods only provide depth-averaged anisotropy and radar only provide anisotropy information in the horizontal plane, perpendicular to the direction of wave propagation. Recent advances in phase coherent radar systems, and data processing for use in inferring ice anisotropy have resulted in an increase in the acquisition of observational fabric data \citep{Dall2010, Young2021b, Young2021, Jordan22, Ershadi2022}. These surveys provide the spatial variability of fabric in more extensive areas and aid understanding of the relation between ice anisotropy and ice flow. Prior to this, the lack of observational data to compare with and constrain models has long hindered the progression of ice fabric modelling.

Although we concentrate only on the coupling of the anisotropy field to the velocity and stress fields in the presented study, we note that advances in a full coupling of crystal orientation tensor evolution with viscosity are ongoing with challenges remaining regarding numerical instability \citep{Gerber2023}. Several modelling studies investigating the effects of an anisotropic rheology on ice flow dynamics at ice divides in two dimensions \citep{Martin2009, Martín2012, Lilien2023} and in ice streams \citep{Gerber2023, Richards2023} have shown the importance of including anisotropy evolution in ice flow models. In idealised, two-dimensional simulations, \cite{Richards2022} investigate the influence of the strain-rate and spin on the fabric type. Building on the work of \cite{Thorsteinsson2003}, \cite{Gillet2006}, and \cite{Pettit2007}, \cite{Rathmann2021} investigated the coupling between anisotropy and viscosity through enhancement factors which result in greater shearing perpendicular to the predominant $c$-axis direction in two-dimensional numerical ice-flow simulations. Using a computationally-efficient anisotropic flow model with a simplified representation of anisotropy coupled to the viscosity via enhancement factors, \cite{McCormack2022} studied the effect of anisotropy on larger-scale three-dimensional models.

The coupling of anisotropy evolution in ice flow models is not without complication due to challenges with numerical stability, the parameter choice within the anisotropy evolution equation, and the coupling with viscosity in the Glen's flow law equation. Furthermore, because an isotropic assumption has traditionally been employed in ice sheet models, model parameters have been optimised to fit isotropic rather than anisotropic ice flow dynamics. For these reasons, a thorough investigation of each step in the process of modelling ice anisotropy is needed.

In this paper, we present the first model of the three-dimensional anisotropy field of an ice rise using the finite element model Elmer/Ice \citep{Gagliardini2013} across $100$ partitions and applied to Derwael Ice Rise. We investigate the dependence of the anisotropy field on the strain rate and deviatoric stress fields without coupling with the viscosity and without re-crystallisation terms, which allows analysis in the absence of the additional feedback complexity. Our simulations provide novel ice fabric predictions for a number of three-dimensional ice-flow settings including areas of vertical-shear-dominated flank flow, grounding zones and shear zones. We highlight areas where geophysical measurements of ice-fabric types would be most informative to constrain relevant model parameters. Furthermore, the direction of the $c$-axis across the ice rise is investigated to identify regions where a vertical $c$-axis assumption is valid in radar measurements of anisotropy, providing a novel method for determining the error in eigenvalues with such an assumption.

\subsection{Motivation}

Our study is motivated by a lack of progress in large-scale ice-sheet models considering crystal orientation fabric. This is despite observations of strong crystal orientation fabric in polar ice \citep{Alley1988}, knowledge of how anisotropic polar ice is \citep{Duval1983} and field evidence of the effect crystal orientation fabric has on ice flow \citep{Gerber2023}. After the introduction of the crystal orientation tensor \citep{godert2003}, the flow-induced crystal orientation fabric evolution can be considered by large-scale ice-sheet models \citep{Gagliardini2013}. 

We believe that the main reasons for the lack of progress are: (a) the numerical implementation of crystal orientation fabric evolution is challenging \citep{Seddik2011}, (b) the interpretation of model output and comparison with non-comprehensive observations is complex \citep{Jordan22}, and (c) essential model parameters within the theory are not yet constrained by observations \citep{Ma2010}.

Numerically, the main issue is that numerical dispersion tends to break down the orientation tensor properties. Here, we present a numerically-robust, three-dimensiontal model, discuss a framework to interpret model output for comparison with observations, and highlight areas where observations could better constrain model parameters.

\subsection{Derwael Ice Rise (DIR)}

\begin{figure}
\includegraphics[width=0.95\textwidth]{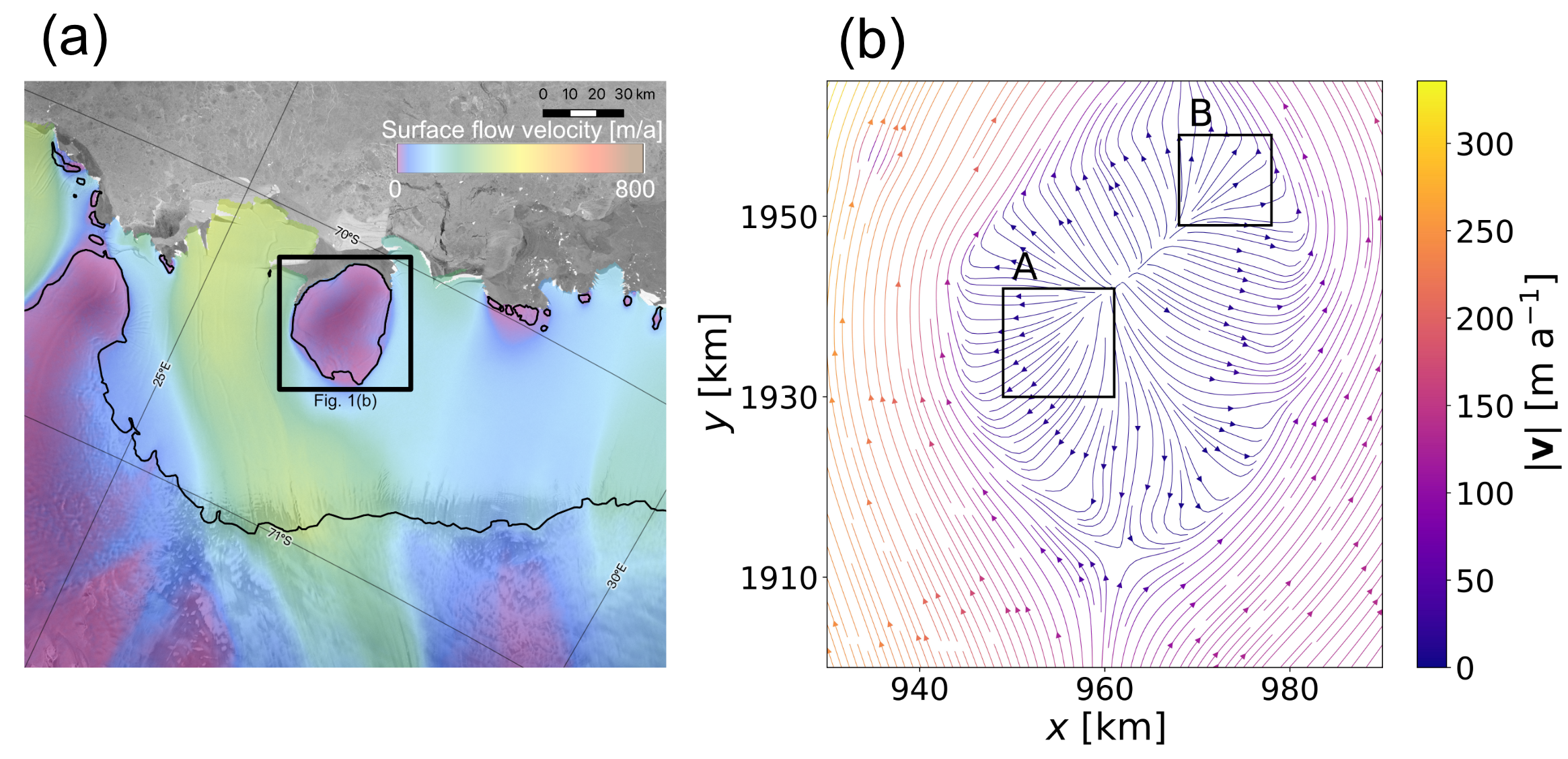}
\caption{In (a), an overview is shown of Derwael Ice Rise, with the surrounding ice shelves and the grounding line \citep{Jezek2013, Rignot2017} The velocity field is shown in colour. In (b), the simulated upper surface velocity field of Derwael Ice Rise is shown, based on simulations in \cite{Henry2023}. The boxes A and B show the areas referred to as the areas of high horizontal divergence at the tails of the flow divide.}
\label{overview}
\end{figure}

Ice rises are an ideal study location for understanding ice-flow processes, because transitions between different flow regimes occur over comparatively short spatial scales. Furthermore, ice rises regulate flow from the Antarctic Ice Sheet towards the open ocean, controlling ice shelf velocities and the continental grounding line position \citep{Favier2015, Schannwell2019, Schannwell2020, Henry2022}. Formation, evolution and disintegration of ice rises occur over glacial-interglacial timescales, meaning that remnants of ice properties such as temperature and anisotropy from previous flow regimes may become stored in the slow flowing ice of an ice rise.

Ice rises typically have clear ice divides transitioning into a flank-flow regime on all sides with little to no basal sliding \citep{Matsuoka2015}. The grounding line is typically located a few tens of kilometres away from the divide, and the flow field transitions to the surrounding ice shelves through narrow shear zones with large horizontal shear strain rates. We chose Derwael Ice Rise (DIR, Fig. \ref{overview}) in East Antarctica as our study site for two reasons. First, we can rely on a predicted three-dimensional steady-state velocity field developed in a previous study \citep{Henry2023}. This velocity field is based on a transient, thermomechanically-coupled full Stokes model with an isotropic rheology. The model was forced with an observationally-constrained surface mass balance field and predictions of the model output were validated with extensive radar observations which are available for this Ice Rise \citep{Koch2023}. Second, previous studies suggest that DIR is likely close to steady state \citep{Callens2016} possibly with a minor amount of thinning \citep{Drews2015}. This justifies the steady-state assumption applied here. DIR is a marine-based ice rise and has a flow divide in the form of a curved arc extended south-west to north-east. The maximum ice thickness is roughly $630$\,m and flow velocities in the flank are typically slower than $10$\,m\,a$^{-1}$.

\section{Methods}

The simulations use output from \cite{Henry2023} as a starting point in which the finite element model Elmer/Ice \citep{Gagliardini2013} was applied to DIR. Most importantly, we use the predicted three-dimensional, steady-state  velocity field to make predictions of the anisotropic ice properties resulting from a one-way coupling with the crystal orientation evolution equations detailed below.

\subsection{Governing equations}

The equations of motion for Stokes flow are written as
\begin{equation}
    \div (\bm{\tau} - P \textbf{I}) + \rho_i \textbf{g} = 0,
\end{equation}
where $\bm{\tau}$ is the deviatoric stress tensor, $P$ is the pressure, $\mathbf{I}$ is the identity matrix, $\rho_i$ is the ice density and $\textbf{g} = g \hat{\mathbf{e}}_z$ is the gravitational acceleration. The ice is subject to an incompressibility condition,
\begin{equation}
    \div \textbf{u} = 0,
\end{equation}
and the Glen's flow law,
\begin{equation}\label{glen}
    \bm{\tau} = 2 \eta \dot{\bm{\varepsilon}},
\end{equation}
which describes the nonlinear dependence between the strain rate tensor, $\dot{\bm{\varepsilon}}$, and the deviatoric stress tensor. The effective viscosity, $\eta$, is
\begin{equation}
    \eta = \frac{1}{2} EA(T')^{-1/n} \dot{\varepsilon}_e^{(1-n)/n}.
\end{equation}
where $E$ is an enhancement factor which is spatially and temporally constant here, $A(T')$ is the ice fluidity which is dependent on the ice temperature relative to the pressure melting point, $T'$, which is solved using a temperature evolution equation as described in \cite{Henry2023}. Here, $n$ is the Glen's flow law exponent, and $\dot{\varepsilon}_e = \sqrt{\tr \dot{\bm{\varepsilon}}^2/2}$ is the effective strain rate.

Keeping the velocity and stress fields constant in time, a semi-Lagrangian anisotropy evolution equation is coupled and simulations are performed for $20000$ years. To initialise the anisotropy simulation, the simulation named \textit{n3E0.5dsdt50} in \cite{Henry2023} is used. The simulation time of $20000$ years was deemed appropriate given that in \cite{Henry2023}, it was predicted that the ice in DIR is roughly $8000$ years old at a depth of $95$ \%.

The anisotropy of ice is described by the second and fourth order orientation tensors, $\mathbf{a}^{(2)}$ and $\mathbf{a}^{(4)}$, respectively defined by 
\begin{equation}
    {a_{ij}}^{(2)} = \langle c_i c_j \rangle
\end{equation}
and
\begin{equation}
    {a_{ijkl}}^{(4)} = \langle c_i c_j c_k c_l \rangle,
\end{equation}
where $\mathbf{c}$ is the $c$-axis unit vector and the operator $\langle \rangle$ denotes the average over all the grains that compose the ice polycrystal. 

The representation of the anisotropy field using a crystal orientation tensor,
\[ \mathbf{a}^{(2)} = \begin{pmatrix}
a^{(2)}_{xx} & a^{(2)}_{xy} & a^{(2)}_{xz} \\
a^{(2)}_{yx} & a^{(2)}_{yy} & a^{(2)}_{yz} \\
a^{(2)}_{zx} & a^{(2)}_{zy} & a^{(2)}_{zz} \end{pmatrix},\]
describing the $c$-axis distribution is ideal for comparison with ice core or radar observations in that an equivalent crystal orientation tensor can be constructed from observational data by determining the cross products of cosines of each $c$-axis and summing up such that
\[ \mathbf{a}^{(2)} = \begin{pmatrix}
a^{(2)}_{xx} & a^{(2)}_{xy} & a^{(2)}_{xz} \\
a^{(2)}_{yx} & a^{(2)}_{yy} & a^{(2)}_{yz} \\
a^{(2)}_{zx} & a^{(2)}_{zy} & a^{(2)}_{zz} \\
\end{pmatrix}
= \frac{1}{N} \begin{pmatrix}
\sum l_i^2 & \sum l_i m_i & \sum l_i n_i \\
\sum m_i l_i & \sum m_i^2 & \sum m_i n_i \\
\sum n_i l_i & \sum n_i m_i & \sum n_i^2 \\
\end{pmatrix}.,\]
where $N$ is the number of $c$-axes being summed over. The cosines of the $c$-axis vectors are defined by taking the cosine of the angle between the $c$-axis vector, $\bm{c}_i$, and each positive coordinate axis, $\bm{\hat{e}}_x$, $\bm{\hat{e}}_y$ and $\bm{\hat{e}}_z$, so that 
\begin{equation}
\begin{aligned}
    l_i & = \frac{\bm{c}_i \cdot \bm{\hat{e}}_x}{|\bm{c}_i|} \\
    m_i & = \frac{\bm{c}_i \cdot \bm{\hat{e}}_y}{|\bm{c}_i|} \\
    n_i & = \frac{\bm{c}_i \cdot \bm{\hat{e}}_x}{|\bm{c}_i|}.
\end{aligned}
\end{equation}
From the orientation tensor, a number of metrics can be calculated to investigate the degree of crystal orientation, the predominant $c$-axes directions and the fabric type. The eigenvalues, namely $\lambda_1$, $\lambda_2$ and $\lambda_3$, and the corresponding eigenvectors, $\mathbf{v}_1$, $\mathbf{v}_2$ and $\mathbf{v}_3$, of the crystal orientation tensor $\mathbf{a}^{(2)}$ satisfy
\begin{equation}
    \mathbf{a}^{(2)} \mathbf{v} = \lambda \mathbf{v}
\end{equation}
defining each eigenvalue by its subscript such that $\lambda_1 \leq \lambda_2 \leq \lambda_3$. In the case of randomly oriented $c$-axes or isotropic ice, all three eigenvalues are similar in size such that $\lambda_1 \approx \lambda_2 \approx \lambda_3 \approx 1/3$. In areas where $\lambda_3 \geq \lambda_1 \approx \lambda_2$, ice fabric is said to have a single maximum fabric with the majority of $c$-axes pointing in a single direction. Where $\lambda_2 \approx \lambda_3 \geq \lambda_2$, ice is said to have a girdle fabric, with $c$-axes following an arc or circle.

The evolution of the second-order orientation tensor can be described by
\begin{equation}
\label{evolution}
\begin{aligned}
    \bigg(\frac{\partial}{\partial t} + \mathbf{u} \cdot \grad \bigg) \mathbf{a}^{(2)} & = \mathbf{W}\mathbf{a}^{(2)} - \mathbf{a}^{(2)}\mathbf{W} \\
     & \phantom{=} - \iota (\mathbf{C}\mathbf{a}^{(2)} + \mathbf{a}^{(2)}\mathbf{C} - 2 \mathbf{a}^{(4)} : \mathbf{C}),
\end{aligned}
\end{equation}
where the tensor $C$ is defined by
\begin{equation}\label{C}
    \mathbf{C} = (1 - \alpha) \dot{\bm{\varepsilon}} + \alpha \frac{1}{2 \eta_0} \bm{\tau},
\end{equation}
and $\alpha$ is the so-called interaction parameter, which describes the relative influence of the strain rate and stress tensors. The parameter $\iota$ determines the rate at which the crystal orientation tensor is influences by the weighted combination of strain-rate and deviatoric stress tensors. The last term, $\alpha \frac{1}{2 \eta_0} \bm{\tau}$, in Eq. (\ref{C}) is represented as $\alpha k_s A \tau_e^{n-1} \bm{\tau}$ in \cite{Gagliardini2013}, but the simulation calculations are equivalent. The spin of the ice is denoted by the spin tensor, $\textbf{W}$, and
\begin{equation}
    \eta_0 = \frac{1}{2} (EA)^{-1/n} \bigg( \frac{1}{2} \text{tr}(\dot{\bm{\varepsilon}}^2)^{\frac{1-n}{2n}} \bigg).
\end{equation}

Early work recognised the influence of the cumulative strain and stress on ice crystal $c$-axis orientation \citep{Alley1988}, which led to the development of a crystal orientation tensor evolution equation (Eq. (\ref{evolution})) dependent on the velocity gradient through the spin and strain rate tensors, and the deviatoric stress tensor \citep{Gillet2006}. Although it is known that the velocity and stress fields have an influence on the anisotropy field of ice, it remains unclear what the relative influence is. The anisotropy evolution equation (Eq. (\ref{evolution})) is made up of various terms, with the spin tensor, $\bm{W}$, acting to rotate the crystal orientation tensor to follow the spin of the ice. Acting opposite to the spin tensor is the tensor, $\bm{C}$, which is a weighted combination of the strain rate tensor, $\bm{\dot{\varepsilon}}$, and the deviatoric stress tensor, $\bm{\tau}$. This combination comes from the fact that the behaviour of macroscopic materials can be limited by two extreme approximations: uniform stress, where the stress in the crystals is assumed to be identical to the macroscopic stress, and Taylor or uniform strain rate, where the strain rate in the crystals is assumed to be identical to the macroscopic strain rate \citep{Gagliardini2009}. The choice of $\alpha$ and $\iota$ in previous studies have been motivated by assumptions of the relative influence of stress and strain rate on crystal orientation evolution. The last term in the equation describes the influence of the higher-order crystal orientation tensor, $\bm{a}^{(4)}$ on the $3 \times 3$ crystal orientation tensor, $\bm{a}^{(2)}$. In reality, $\bm{a}^{(2)}$ is dependent on further, higher-order tensors, but it has been shown that this dependence is negligible \citep{Chung2002}.

To solve Eq. (\ref{evolution}) we also require a relation between $\mathbf{a}^{(2)}$ and $\mathbf{a}^{(4)}$, a closure approximation, and we use an invariant-based optimal fitting closure approximation \citep{Gillet2006}. The distribution of $c$-axes can be described by tensors of ever-increasing order, but a compromise is found by using the closure approximation in order to reduce computation time. The effect of the choice of $\alpha$ and $\iota$ on the anisotropy field is investigated using combinations of parameters from previous literature presented in Table \ref{parameters}.

\begin{table}\centering
\caption{Parameter combinations}
\label{parameters}
\begin{tabular}{l c c c c}\hline
 & $\alpha$ & $\iota$ & Source \\
\hline
(a) & $0$ & $1$ & \cite{Martin2009} \\
(b) & $0$ & $0.6$ & \cite{Seddik2011} \\
(c) & $1$ & $1$ & \cite{Martín2012} \\
(d) & $0.06$ & $1$ & \cite{Gagliardini2013} \\
\end{tabular}
\end{table}

The value of $\alpha$ determines the relative influence of the strain rate and deviatoric stress tensors on anisotropy evolution. For example, a value of $\alpha=0$, means that there is dependence on the strain rate tensor but no dependence on the deviatoric stress. Alternatively, a value of $\alpha = 1$ provides dependence on the deviatoric stress tensor but no dependence on the strain rate tensor. The parameter $\iota$ adjusts the rate at which the anisotropy field develops in response to the weighted combination of the strain rate and deviatoric stress fields and takes a value between $\iota=0$ and $\iota=1$.

Initially, all ice in the model domain is isotropic, described by the tensor $\mathbf{a}^{(2)} = \frac{1}{3} \textbf{I}$, where $\textbf{I}$ is the $3\times3$ identity matrix. During transient simulation of the anisotropy field dependent on the velocity field, the upper surface is assigned an isotropic boundary conditions on all ice entering the domain due to accumulation.

The anisotropic evolution model described in Eq. (\ref{evolution}) together with its boundary and initial conditions is solved using a semi-Lagrangian method as described in \cite{Martin2009}. The determinant of $\mathbf{a}^{(2)}$ gives information about the degree of crystal orientation of the fabric \citep{Advani1987}. We define here the degree of crystal orientation with the scalar value,
\begin{equation}
    d=1-3^3 \det (\mathbf{a}^{(2)} ).
\end{equation}
The degree of crystal orientation varies from zero for isotropic ice to one for single-maximum fabric, and is independent of the frame of reference as the determinant is an invariant. We constrain Equation (\ref{evolution}) with the condition,
\begin{equation}
    \bigg(\frac{\partial}{\partial t} + \mathbf{u} \cdot \grad \bigg) d \geq 0,
\end{equation}
meaning that ice is constrained to increase in degree of crystal orientation over time. This constraint on the degree of crystal orientation stops numerical dispersion breaking the simulation in areas with high strain rates. Our method is also straightforward to implement in large scale models that run in parallel environments.

\section{Results}

\begin{figure*}
\centering{\includegraphics[width=0.95\textwidth]{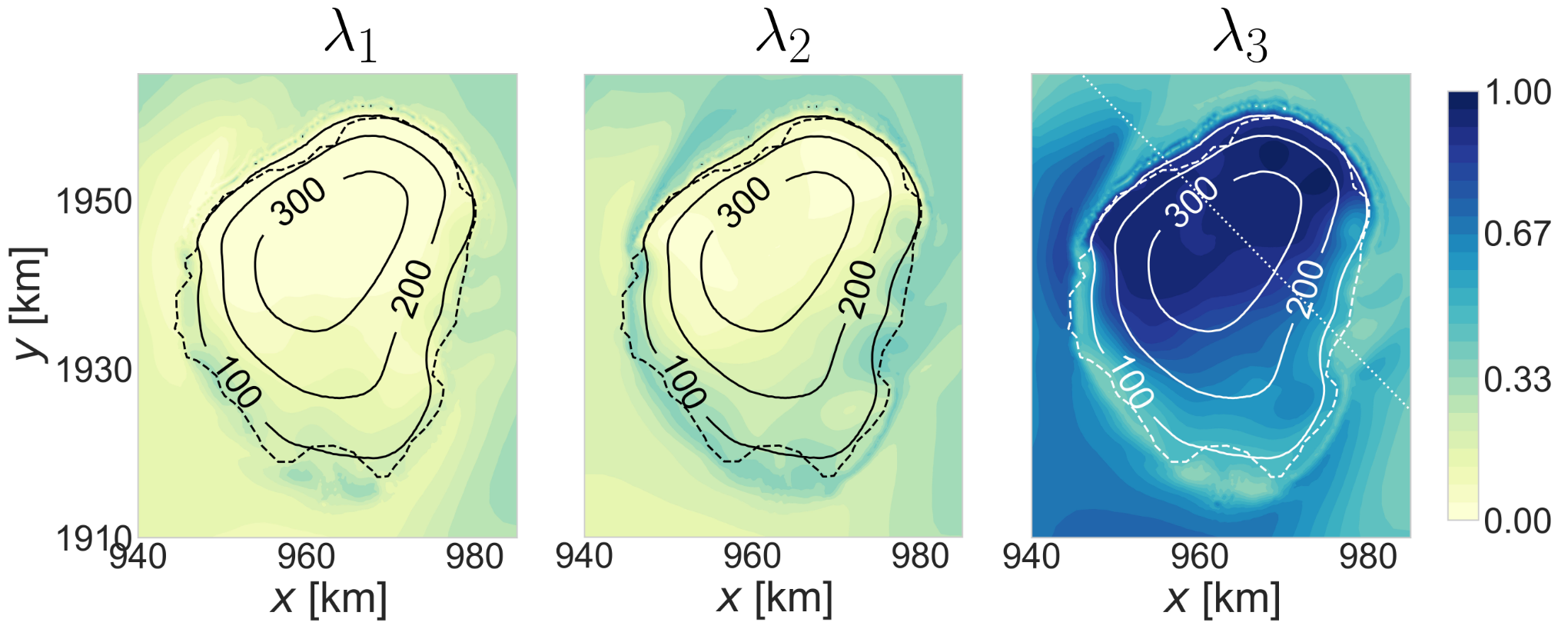}}
\caption{The eigenvalues, $\lambda_1$, $\lambda_2$ and $\lambda_3$, of the crystal orientation tensor in the $\alpha=0$, $\iota=1$ simulation at an elevation of $z=0$, corresponding to sea level. The solid lines, black in the plots showing $\lambda_1$ and $\lambda_2$, and white in the plot showing $\lambda_3$, are contours of depth below the upper ice surface and the dashed lines show the grounding line. The dotted line in the $\lambda_3$ figure shows where the cross-section in Fig. \ref{CS_lambda3} is taken.}
\label{lambdas}
\end{figure*}

\begin{figure*}
\center\noindent\includegraphics[width=0.9\textwidth]{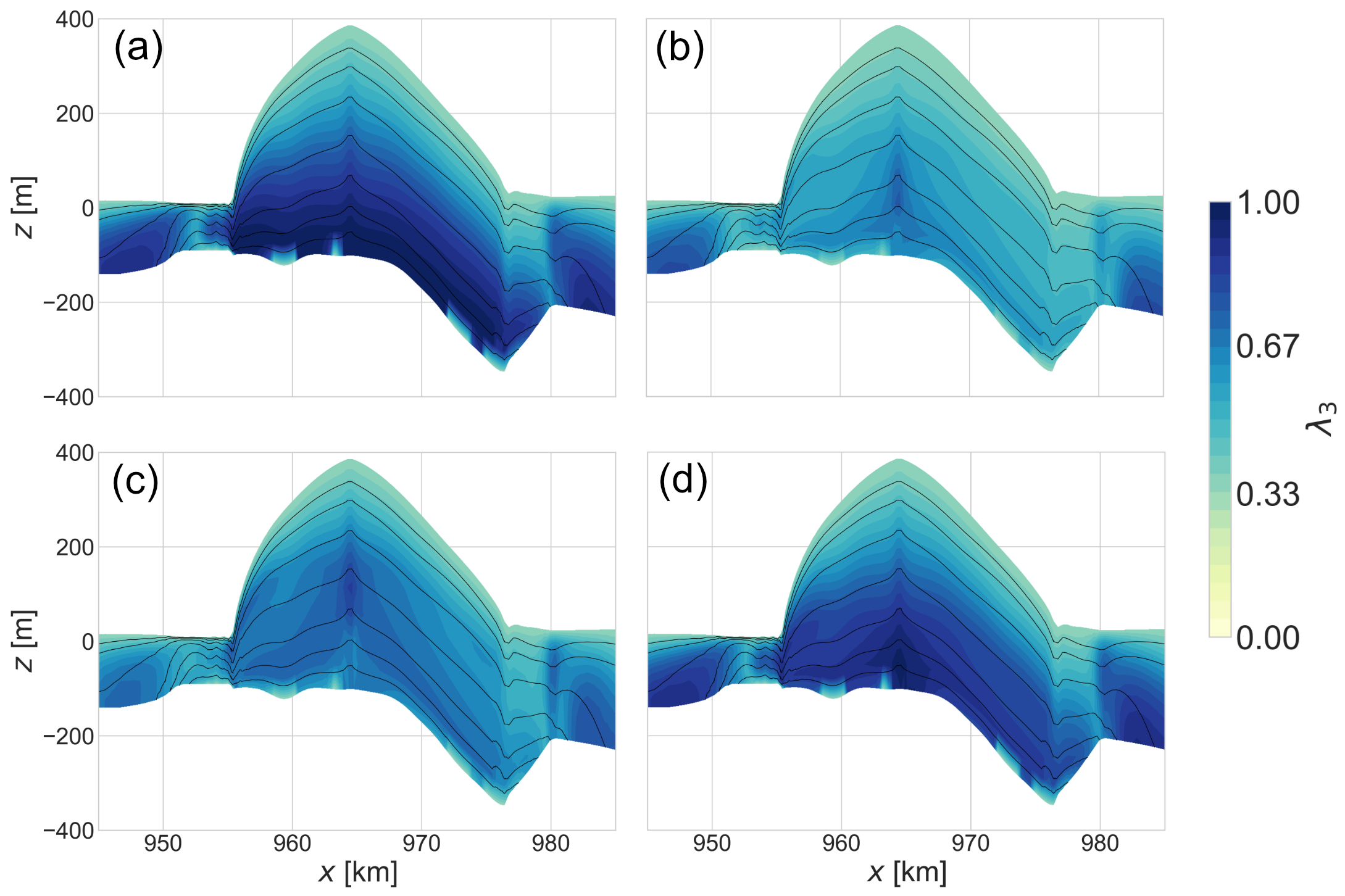}
\caption{Cross-sections through the flow divide as shown in the Fig. \ref{lambdas} showing $\lambda_3$, the largest eigenvalue for (a) the $\alpha=0$, $\iota=1$ simulation, (b) the $\alpha=0$, $\iota=0.6$ simulation, (c) the $\alpha=1$, $\iota=1$ simulation and (d) the $\alpha=0.06$, $\iota=1$ simulation. The solid lines show isochrones.}
\label{CS_lambda3}
\end{figure*}

\subsection{Analysis of simulated anisotropy field}

\begin{figure}
\center\noindent\includegraphics[width=0.6\textwidth]{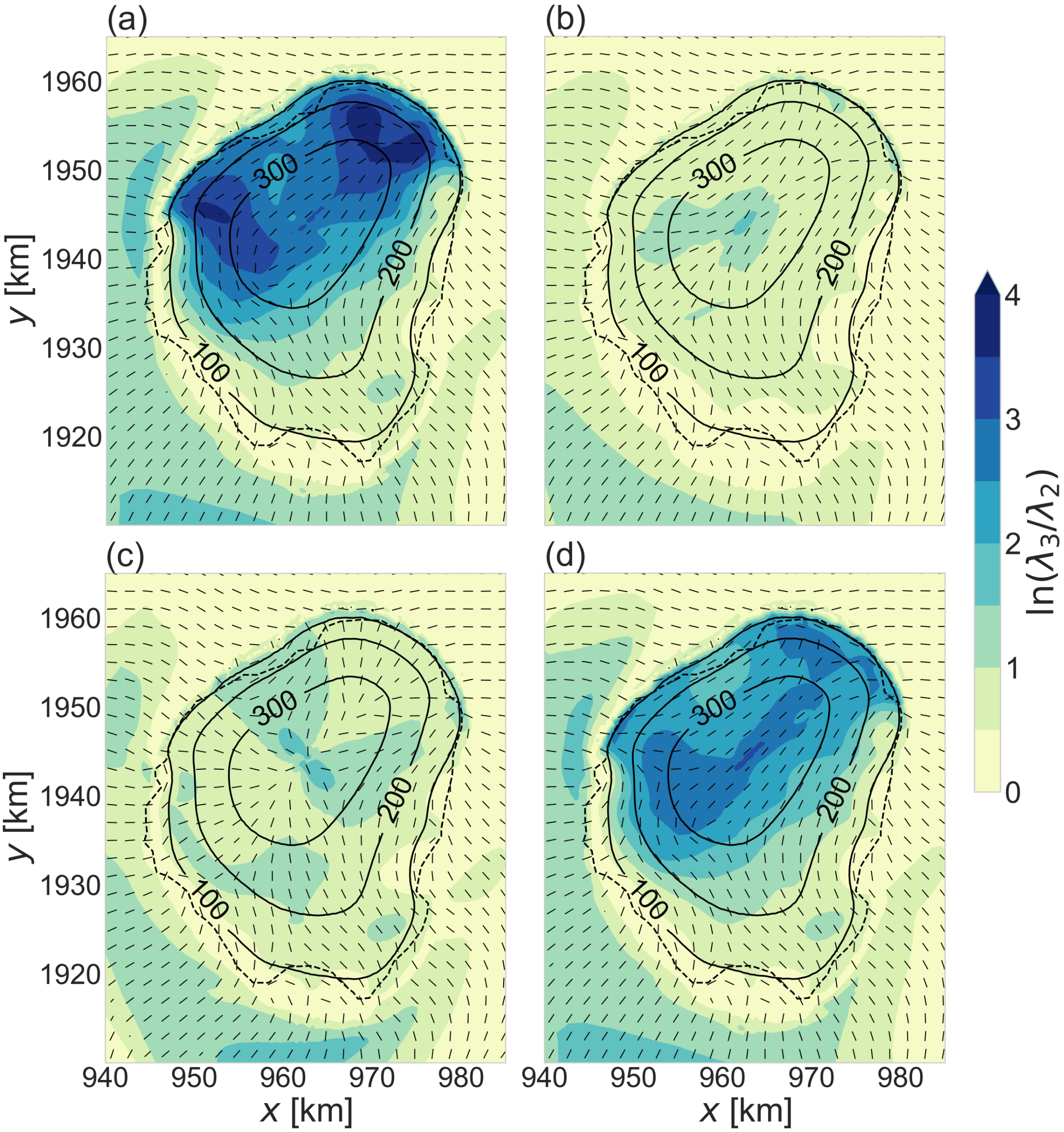}
\caption{The ratio of the two larger eigenvalues of the $3 \times 3$ crystal orientation tensor at $z=0$ for (a) the $\alpha=0$, $\iota=1$ simulation, (b) the $\alpha=0$, $\iota=0.6$ simulation, (c) the $\alpha=1$, $\iota=1$ simulation and (d) the $\alpha=0.06$, $\iota=1$ simulation. The dashes show the maximum horizontal anisotropy direction. The dashed line shows the grounding line and the solid line show contours of the depth below the upper ice surface.}
\label{lambda_3_2}
\end{figure}

We display predicted eigenvalues for the $\alpha=0$, $\iota=1$ simulation at an elevation of $z=0$ which encompasses an ice-depth range of close to $0$ m in the ice shelf to over $300$ m at the ice rise flow divide where stresses are dominated by vertical compression and lateral extension (Fig. \ref{lambdas}). The main characteristic, which is evident in all simulations, is that the ice fabric evolves from an isotropic material at the surface and under deformation, develops into an anisotropic fabric varying from a single maximum to a girdle fabric (Fig. \ref{CS_lambda3}). In the ice rise interior, the largest eigenvalue, $\lambda_3$, is much larger than $\lambda_1$ and $\lambda_2$, particularly at and surrounding the flow divide (Figs. \ref{overview}b and \ref{lambdas}). This results in greater fabric anisotropy here. Further south in the ice rise, where ice flow is hindered by the convergence of flow between the ice rise and the ice shelf, differences are not as substantial. In the ice rise interior away from the grounding line, differences between the smaller two eigenvalues, $\lambda_1$ and $\lambda_2$ are mostly small, increasing slightly on the stoss side of the ice rise. The stoss side is defined as the side of the ice rise with a flow direction opposing the upstream ice shelf flow direction (Figs. \ref{overview}b and \ref{lambdas}). At the grounding zone on the stoss side of the ice rise, where horizontal convergence of flow occurs, the three eigenvalues are similar in magnitude, whereas at the transition from grounded to floating ice on the lee side of the ice rise, where flow is dominated be extension, a slightly larger difference is seen between the smaller two eigenvalues, $\lambda_1$ and $\lambda_2$. In the ice shelves away from the grounding zone, there are differing patterns in the relative magnitude of the three eigenvalues. In the ice shelf west of the ice rise where velocities are higher than east of the ice rise and extension dominates, the largest eigenvalue is significantly larger in magnitude compared with the two smaller eigenvalues. In the ice shelf to the east of the ice rise, where extension of flow is an active process, differences in magnitude are generally much lower. The ice shelf south of the ice rise, on the stoss side, shows one eigenvalue much larger than the other two, similar to the magnitudes in the ice shelf on the western side of the ice rise.

\subsection{Comparison of simulations with differing parameter choice}

In order to compare simulation results across the four combinations of the parameters $\alpha$ and $\iota$ in Eq. (\ref{evolution}) used in the previous studies stated in Table \ref{parameters}, we use a number of metrics to understand the fabric types evolving in the various flow regimes at Derwael Ice Rise. First, we describe the differences in relative magnitude of the eigenvalues of the crystal orientation tensor across the four simulations. If the logarithmic ratio \citep{Woodcock1977}, $\ln (\lambda_3 / \lambda_2)$, is large, then the $c$-axes are more concentrated in a single direction. The direction of the largest horizontal crystal orientation eigenvector, $\bm{v}_{2, H}$, shows the predominant horizontal $c$-axis direction. 

The simulations with parameter choices of $\alpha=0$, $\iota=1$ (Fig. \ref{lambda_3_2}a) and $\alpha=0.06$, $\iota=1$ (Fig. \ref{lambda_3_2}d) have comparable results, with a much greater $\ln (\lambda_3 / \lambda_2)$ than predicted by the other two simulations. The simulation results differ, however, in the spatial variation of $\ln (\lambda_3 / \lambda_2)$. The $\alpha=0$, $\iota=1$ simulation shows larger values of $\ln (\lambda_3 / \lambda_2)$ at the tails of the ice rise flow divide (Boxes A and B in Fig. \ref{overview}b), whereas the $\alpha=0.06$, $\iota=1$ simulation shows slightly greater relative values along the flow divide. Although not negligible, the simulations with parameter choices of $\alpha=0$, $\iota=0.6$ (Fig. \ref{lambda_3_2}b) and $\alpha=1$, $\iota=1$ (Fig. \ref{lambda_3_2}c) show much lower ratios between the largest and second largest eigenvalues, with the $\alpha=0$, $\iota=0.6$ simulation showing a slightly higher value at the centre of the flow divide. In the $\alpha=1$, $\iota=1$ simulation, a differing pattern of horizontal eigenvectors originating at a point source at the flow divide, whereas the other simulations show alignment of the eigenvector direction along the flow divide. Of note is also the differing eigenvector directions at the tails of the  flow divide, with differing patterns of vector divergence and convergence.

\begin{figure}
\center\noindent\includegraphics[width=0.6\textwidth]{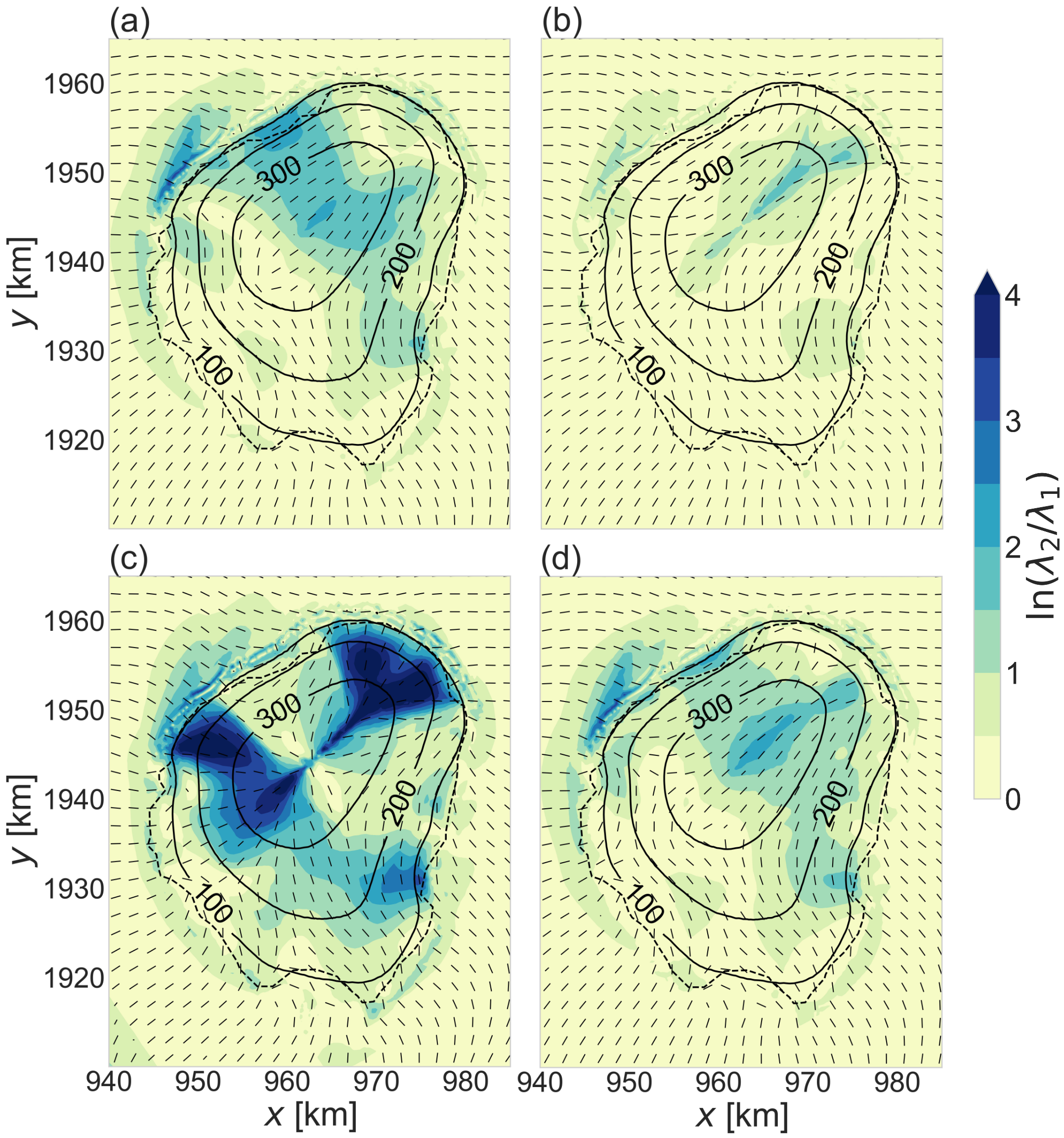}
\caption{The ratio of the two smaller eigenvalues of the $3 \times 3$ crystal orientation tensor at $z=0$ for (a) the $\alpha=0$, $\iota=1$ simulation, (b) the $\alpha=0$, $\iota=0.6$ simulation, (c) the $\alpha=1$, $\iota=1$ simulation and (d) the $\alpha=0.06$, $\iota=1$ simulation. The dashes show the maximum horizontal anisotropy direction. The dashed line shows the grounding line and the solid line show contours of the depth below the upper ice surface.}
\label{lambda_2_1}
\end{figure}

\begin{figure}
\center\noindent\includegraphics[width=0.6\textwidth]{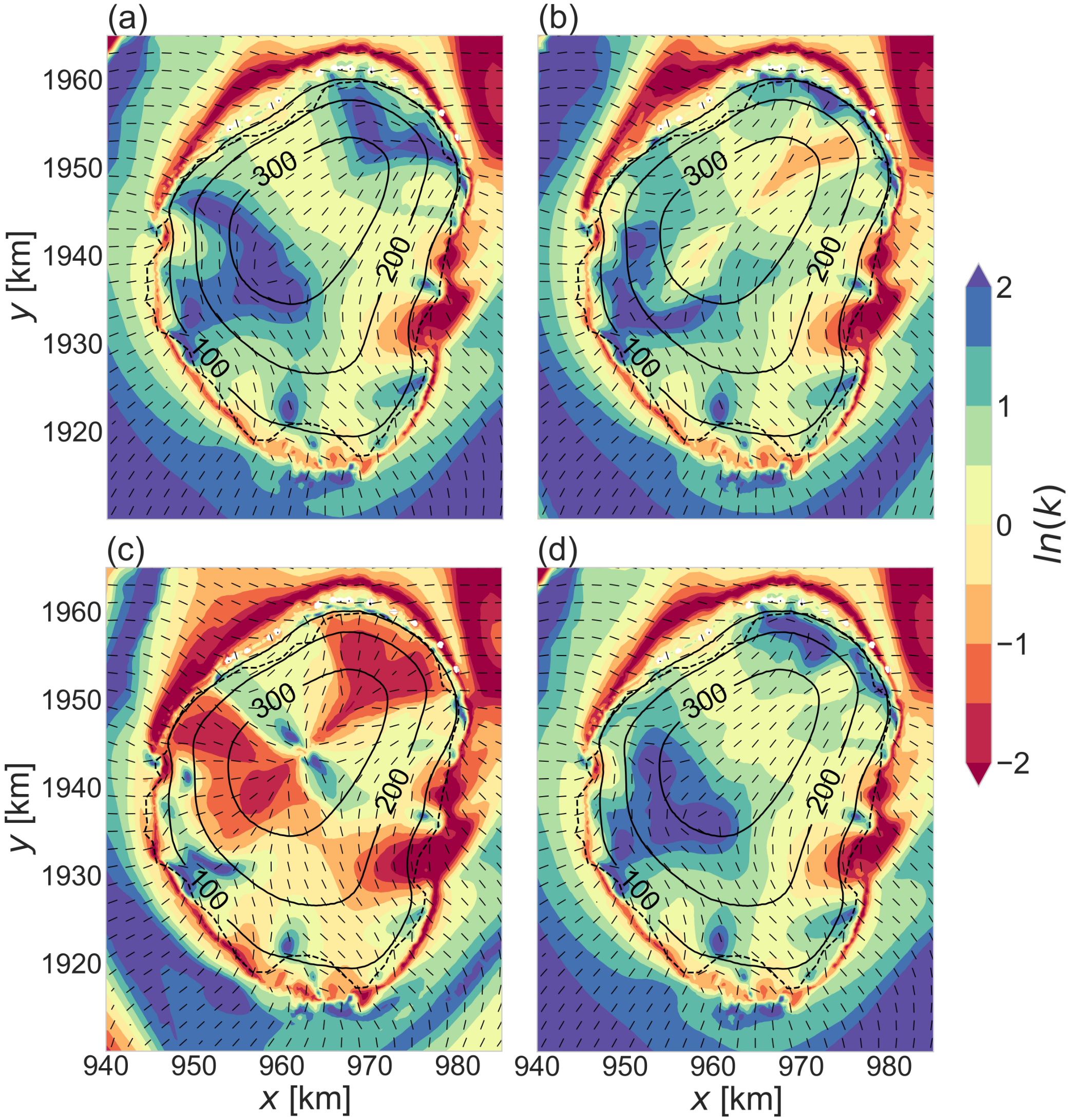}
\caption{The logarithm of the Woodcock $k$ value at an elevation of $z=0$ for (a) the $\alpha=0$, $\iota=1$ simulation, (b) the $\alpha=0$, $\iota=0.6$ simulation, (c) the $\alpha=1$, $\iota=1$ simulation and (d) the $\alpha=0.06$, $\iota=1$ simulation. The dashes show the maximum horizontal anisotropy direction. The dashed line shows the grounding line and the solid line show contours of the depth below the upper ice surface.}
\label{lnk}
\end{figure}

The two smaller eigenvalues, $\lambda_1$ and $\lambda_2$ are investigated using the metric $\ln (\lambda_2 / \lambda_1)$ (Fig. \ref{lambda_2_1}). Large ratios between the two smaller eigenvalues indicates that $c$-axis directions for a given volume of ice are concentrated along an arc. In all simulations, there are high values of $\ln (\lambda_2 / \lambda_1)$ at the flow divide and generally low values in the vicinity of the grounding zone on the stoss side of the ice rise, with values differing between simulations elsewhere. The $\alpha=0$, $\iota=0.6$ (Fig. \ref{lambda_2_1}b) simulation show the lowest values overall. The simulations with parameter choices of $\alpha=0$, $\iota=1$ and $\alpha=0.06$, $\iota=1$ show relatively similar results, with the highest eigenvalue ratio at the flow divide and in the areas of the ice rise perpendicular to the flow divide. The $\alpha=1$, $\iota=1$ simulation (Fig. \ref{lambda_2_1}c) has areas of a large ratio between $\lambda_2$ and $\lambda_1$, much larger than any other simulation and highlights the effect of a strong dependence on the strain-rate tensor on the crystal orientation tensor evolution. Rather than having high values perpendicular to the flow divide, the highest values are along a small band following the flow divide and in the areas of high horizontal divergence at the tails of the flow divide, extending in some areas as far as the grounding line. There are no large differences between $\lambda_2$ and $\lambda_1$ in the ice shelves, with the largest values being concentrated north-west of the ice rise in all simulations.

The Woodcock $k$ value \citep{Woodcock1977}, defined as
\begin{equation}
    k = \frac{\ln (\lambda_3 / \lambda_2)}{\ln (\lambda_2 / \lambda_1)},
\end{equation}
provides a metric by which to investigate which areas are characterised by a single maximum fabric and which are characterised by girdle fabric. Note that $k$ can have values between $k = 0$ and $k = \infty$. In order to better investigate small values of $k$, we plot $\ln(k)$ for each simulation in Table \ref{parameters}. If $\ln(k) > 0$, then the ice is defined as having a single maximum fabric and if $\ln(k) < 0$, the ice is defined as having a girdle fabric. Furthermore, if $\ln(k) >> 0$ and $\ln (\lambda_3 / \lambda_2) >> 0$, then the ice has a strong single maximum. If, on the other hand, $\ln(k) << 0$ and $\ln (\lambda_2 / \lambda_1) >> 0$, then ice has a strong girdle fabric.

In the $\alpha=0$, $\iota=1$ simulation, there are relatively high $\ln(k)$ values at the tails of the flow divide (Fig. \ref{lnk}a), extending in some areas almost to the grounding line. Elsewhere on the ice rise, values of $\ln(k)$ are generally close to zero. The $\alpha=0.06$, $\iota=1$ simulation (Fig. \ref{lnk}d) shows similar results except at the north-east of the ice rise, where high values of $\ln(k)$ are concentrated closer to the grounding line. The $\alpha=0$, $\iota=0.6$ simulation (Fig. \ref{lnk}b) shows higher values of $\ln(k)$ even further from the flow divide and a small area with negative $\ln(k)$ values at the north-eastern end of the flow divide. The $\alpha=1$, $\iota=1$ simulation (Fig. \ref{lnk}c) shows, by far, the most negative values of $\ln(k)$, concentrated at the two tails of the flow divide and small areas with positive values of $\ln(k)$ perpendicular to the flow divide. All four simulations show an almost-continuous band of negative values of $\ln(k)$ at the grounding line or a small distance away from the grounding line in the ice shelf. Moving away from the grounding line, a general increase in values of $\ln(k)$ is seen, with some exceptions.

\subsection{Metrics for comparison with radar data}

In quad-polarimetric radar processing, it is often assumed that because of the dominance of vertical compression, one eigenvector aligns with the vertical direction. In areas where this assumption holds, signal processing can be simplified \citep{Jordan22, Ershadi2022}. Obliquely oriented fabric types can, in theory, also be detected \citep{Matsuoka2009, Rathmann2022b}, but thus far this has not been done for observations which are typically only collected in a nadir-viewing geometry. Here, we present results evaluating the applicability of the assumption of one vertical eigenvector across all four simulations.

\begin{figure}
\center\noindent\includegraphics[width=0.6\textwidth]{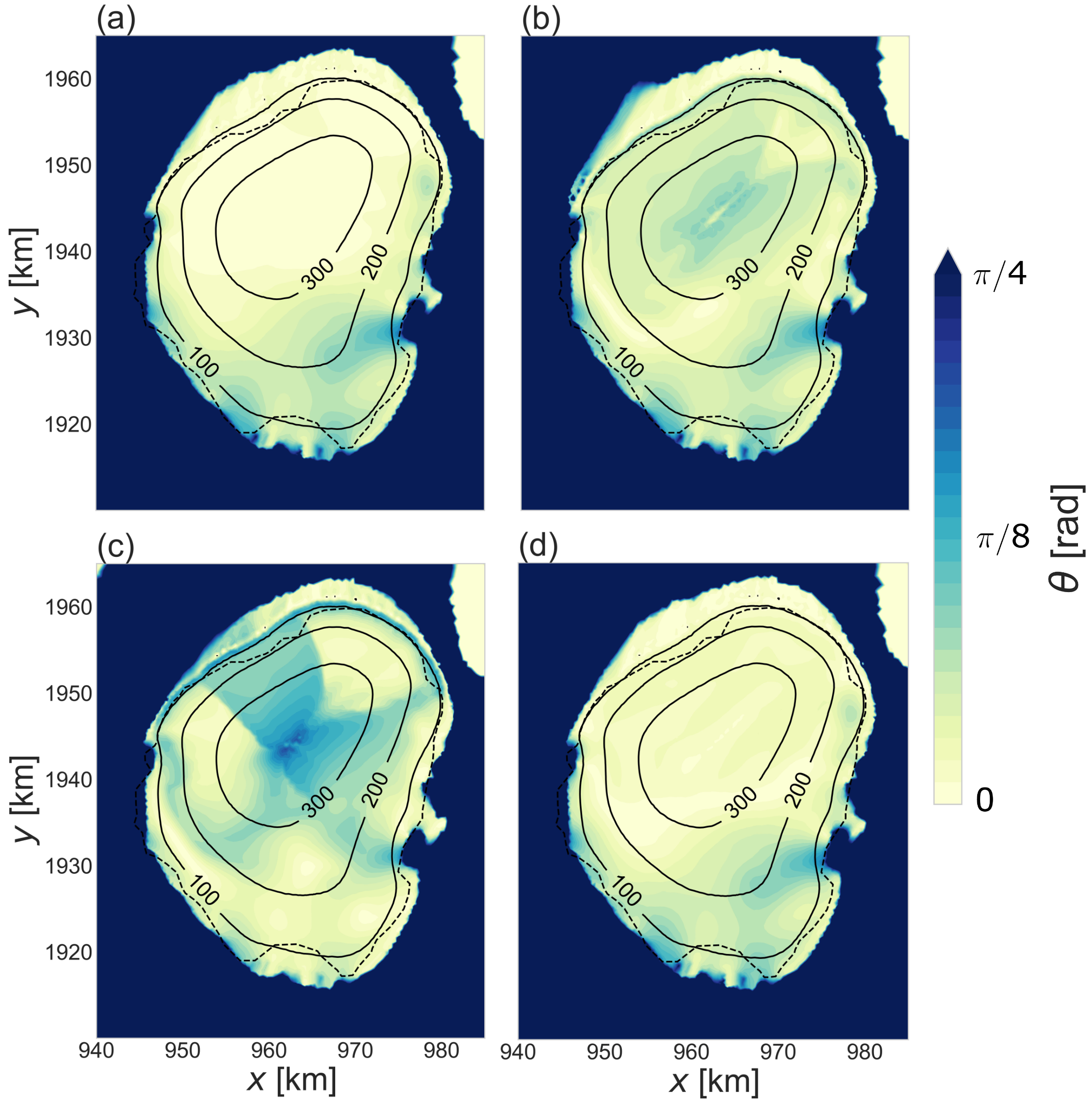}
\caption{The angle between the eigenvector corresponding to the largest eigenvalue and the vertical direction at an elevation of $z = 0$ corresponding to sea level for (a) the $\alpha = 0$, $\iota = 1$ simulation, (b) the $\alpha = 0$, $\iota = 0.6$ simulation, (c) the $\alpha = 1$, $\iota = 1$ simulation and (d) the $\alpha = 0.06$, $\iota = 1$ simulation. The dashed line shows the grounding line and the solid lines show contours for the depth below the surface in metres.}
\label{angle_vert}
\end{figure}

In general, the predicted tilt angle is small in the grounded area across all simulations(Fig. \ref{angle_vert}). Note that the grounded area is the area within the dashed line marking the grounding line. The angle remains small, with similar spatial patterns in the simulations with parameter choices of $\alpha=0$, $\iota=1$ and $\alpha=0.06$, $\iota=1$ (Figs. \ref{angle_vert}a and \ref{angle_vert}d, respectively). In the $\alpha=0$, $\iota=0.6$ simulation (Fig. \ref{angle_vert}b), differences between the $z$-direction and $\bm{v}_3$ are generally small, with the exception of slightly higher values on either side of the flow divide. Spatial patterns on the stoss side of the ice rise are similar to those in the above-mentioned $\alpha=0$, $\iota=1$ and $\alpha=0.06$, $\iota=1$ simulations. The simulation with the largest tilt angles is the $\alpha=1$, $\iota=1$ simulation (Fig. \ref{angle_vert}c), with differences of $\pi/8$ to $\pi/4$ radians at and perpendicular to the flow divide. In all simulations, the tilt angle in the ice shelf is larger a short distance away from the grounding line, except for small, isolated areas.

We furthermore calculate the eigenvectors and eigenvalues of the horizontal crystal orientation tensor, i.e. the upper left $2 \times 2$ part of the $3 \times 3$ tensor $\mathbf{a}^{(2)}$. The reason for this is that if the $3 \times 3$ crystal orientation tensor has one strictly vertical eigenvector, the eigenvalues and eigenvectors of the $2 \times 2$ tensor correspond to the horizontal eigenvalues and eigenvectors of the $3 \times 3$ tensor. We denote the eigenvalues of the $2 \times 2$ tensor by $\lambda_{1,H}$ and $\lambda_{2,H}$, and the corresponding eigenvectors by $\mathbf{v}_{1, H}$ and  $\mathbf{v}_{2,H}$, respectively. By comparing the eigenvalues of the $2 \times 2$, horizontal crystal orientation tensor with the eigenvalues of the $3 \times 3$ crystal orientation tensor, an error estimate can be found for assuming that the eigenvector corresponding to the largest eigenvalue is aligned with the $z$-axis.

\begin{figure}
\center\noindent\includegraphics[width=0.6\textwidth]{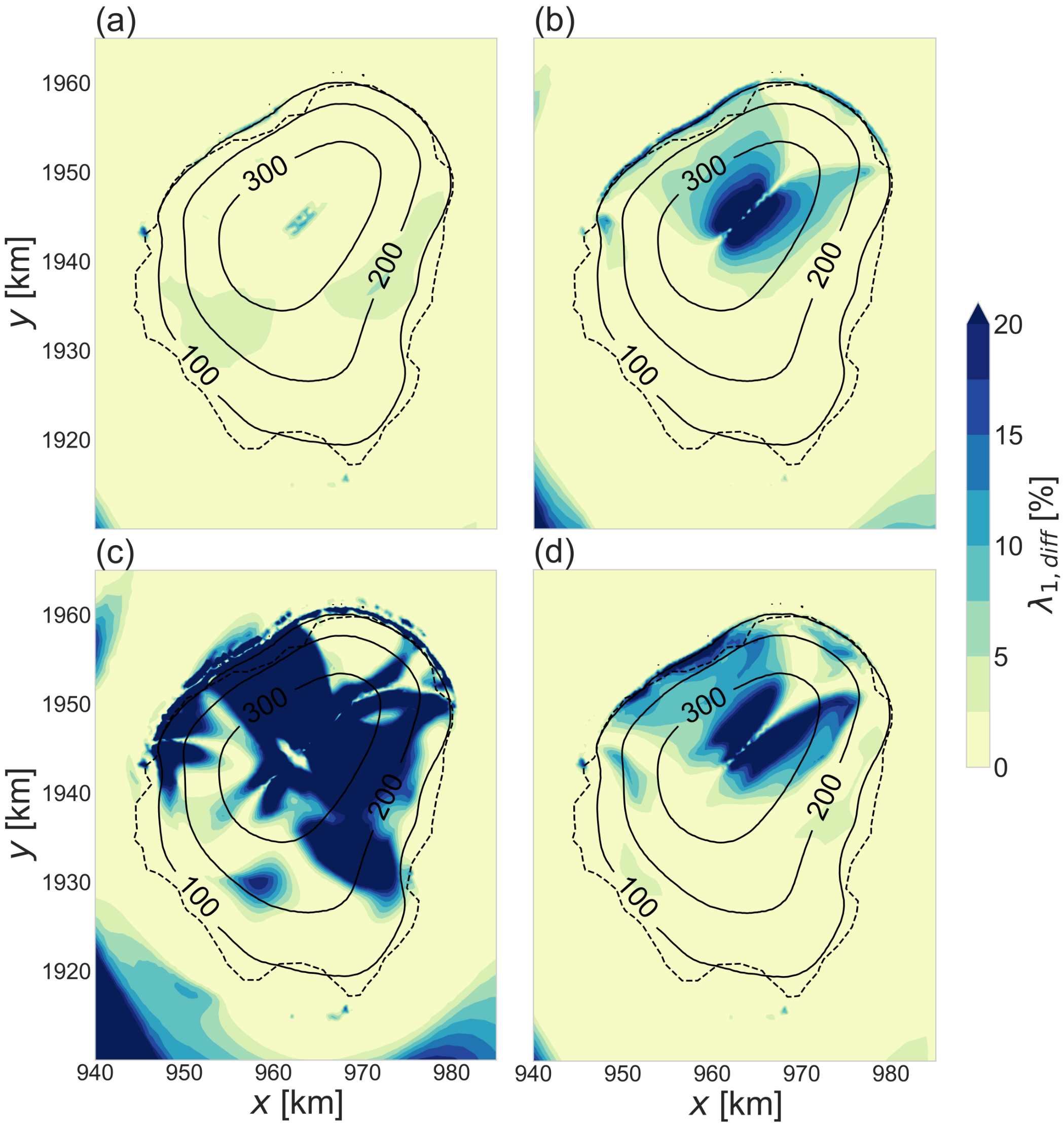}
\caption{The percentage difference between $\lambda_{1,H}$ and $\lambda_1$ at an elevation of $z=0$ for (a) the $\alpha = 0$, $\iota = 1$ simulation, (b) the $\alpha = 0$, $\iota = 0.6$ simulation, (c) the $\alpha = 1$, $\iota = 1$ simulation and (d) the $\alpha = 0.06$, $\iota = 1$ simulation. The solid line contours show depth below surface and the dashed line is the grounding line.}
\label{lambda1_diff}
\end{figure}

\begin{figure}
\center\noindent\includegraphics[width=0.6\textwidth]{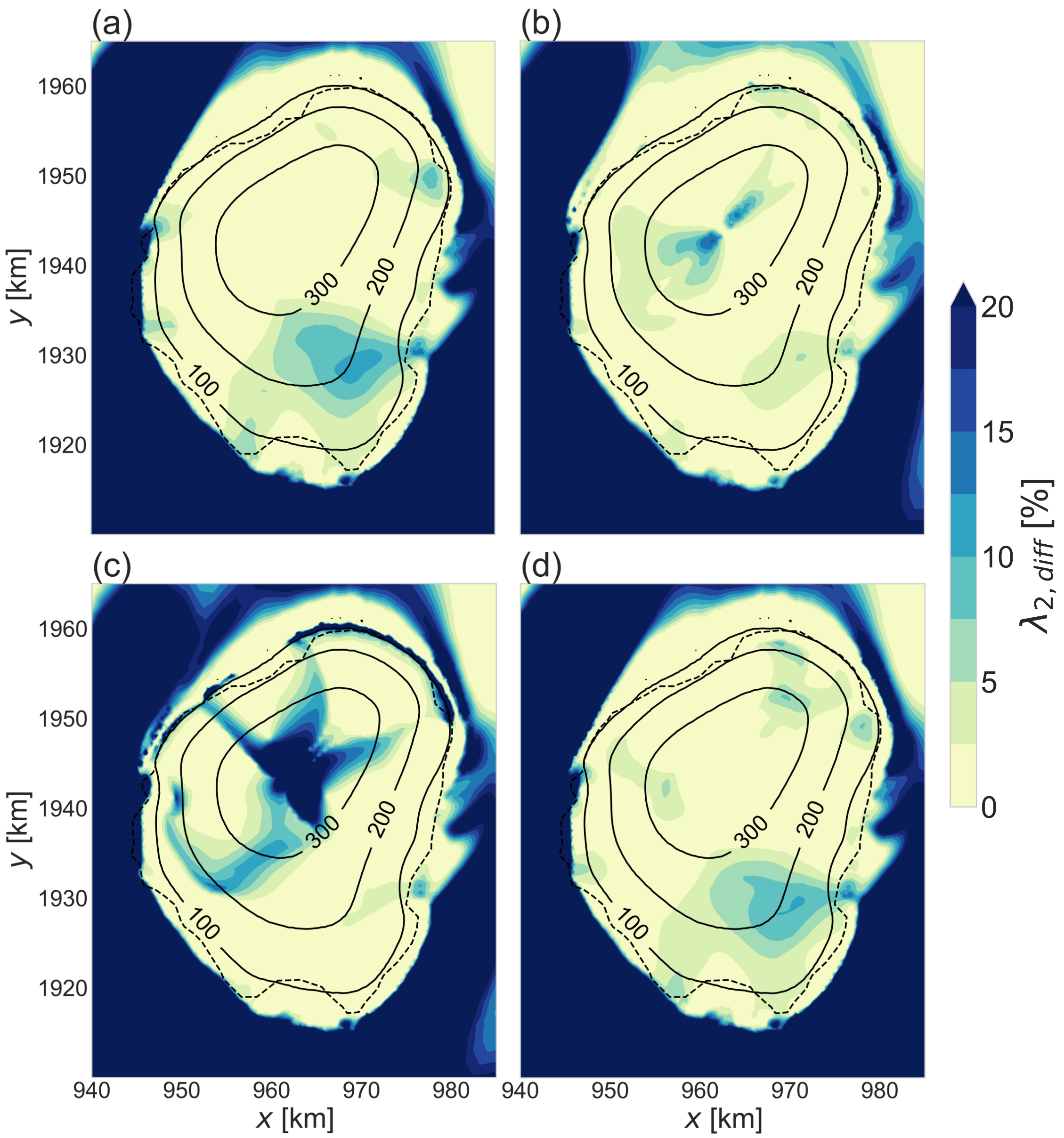}
\caption{The percentage difference between $\lambda_{2,H}$ and $\lambda_2$ at an elevation of $z=0$ for (a) the $\alpha = 0$, $\iota = 1$ simulation, (b) the $\alpha = 0$, $\iota = 0.6$ simulation, (c) the $\alpha = 1$, $\iota = 1$ simulation and (d) the $\alpha = 0.06$, $\iota = 1$ simulation. The solid line contours show depth below surface and the dashed line is the grounding line.}
\label{lambda2_diff}
\end{figure}

We present percentage differences between $\lambda_1$ and $\lambda_{1,H}$ (Fig. \ref{lambda1_diff}), as well as $\lambda_2$ and $\lambda_{2,H}$ (Fig. \ref{lambda2_diff}). We find that if a difference exists between $\lambda_1$ and $\lambda_{1,H}$, or $\lambda_2$ and $\lambda_{2,H}$, then the $2 \times 2$ tensor eigenvalues always underestimate the smaller two $3 \times 3$ tensor eigenvalues. Furthermore, percentage differences tend to be higher for the smaller eigenvalues $\lambda_1$ and $\lambda_{1,H}$ than for $\lambda_2$ and $\lambda_{2,H}$, but exceptions to this are seen. The $\alpha=0$, $\iota=1$ simulation shows a negligible percentage difference between $\lambda_1$ and $\lambda_{1,H}$ (Fig. \ref{lambda1_diff}a) and small percentage differences for $\lambda_2$ and $\lambda_{2,H}$ of up to $~10$ \% (Fig. \ref{lambda2_diff}a). A similar spatial distribution of percentage differences appears for $\lambda_2$ and $\lambda_{2,H}$ in the $\alpha=0.06$, $\iota=1$ simulation (Fig. \ref{lambda2_diff}d). Most notable is the percentage difference between $\lambda_1$ and $\lambda_{1,H}$ in the $\alpha=1$, $\iota=1$ simulation (Fig. \ref{lambda1_diff}c), where values are above $20$ \% in a large area at and perpendicular to the flow divide. In this simulation ($\alpha=1$, $\iota=1$), percentage differences between $\lambda_2$ and $\lambda_{2,H}$ (Figs. \ref{lambda2_diff}c) are high compared to other simulations, but the high values localised at and in the area perpendicular to the flow divide occupy a smaller area than the percentage differences between $\lambda_1$ and $\lambda_{1,H}$ (Fig. \ref{lambda1_diff}c). In the $\alpha=0$, $\iota=0.6$ and the $\alpha=0.06$, $\iota=1$ simulations, percentage differences between $\lambda_1$ and $\lambda_{1,H}$ are negligible across a large portion of the ice rise, in particular on the stoss side (Figs. \ref{lambda1_diff}b,d). The highest percentage differences are either side of the flow divide with values up to and over $20$ \%, reducing further away from the flow divide. Percentage differences between $\lambda_2$ and $\lambda_{2,H}$ in the $\alpha=0$, $\iota=0.6$ simulation are negligible except for small areas at the flow divide (Fig. \ref{lambda2_diff}b). As the assumption that the eigenvector corresponding with the largest eigenvalue is made for grounded ice, we do not analyse the differences between eigenvalues of the $2 \times 2$ and the $3 \times 3$ crystal orientation tensors in the ice shelves.

\begin{figure}
\center\noindent\includegraphics[width=0.6\textwidth]{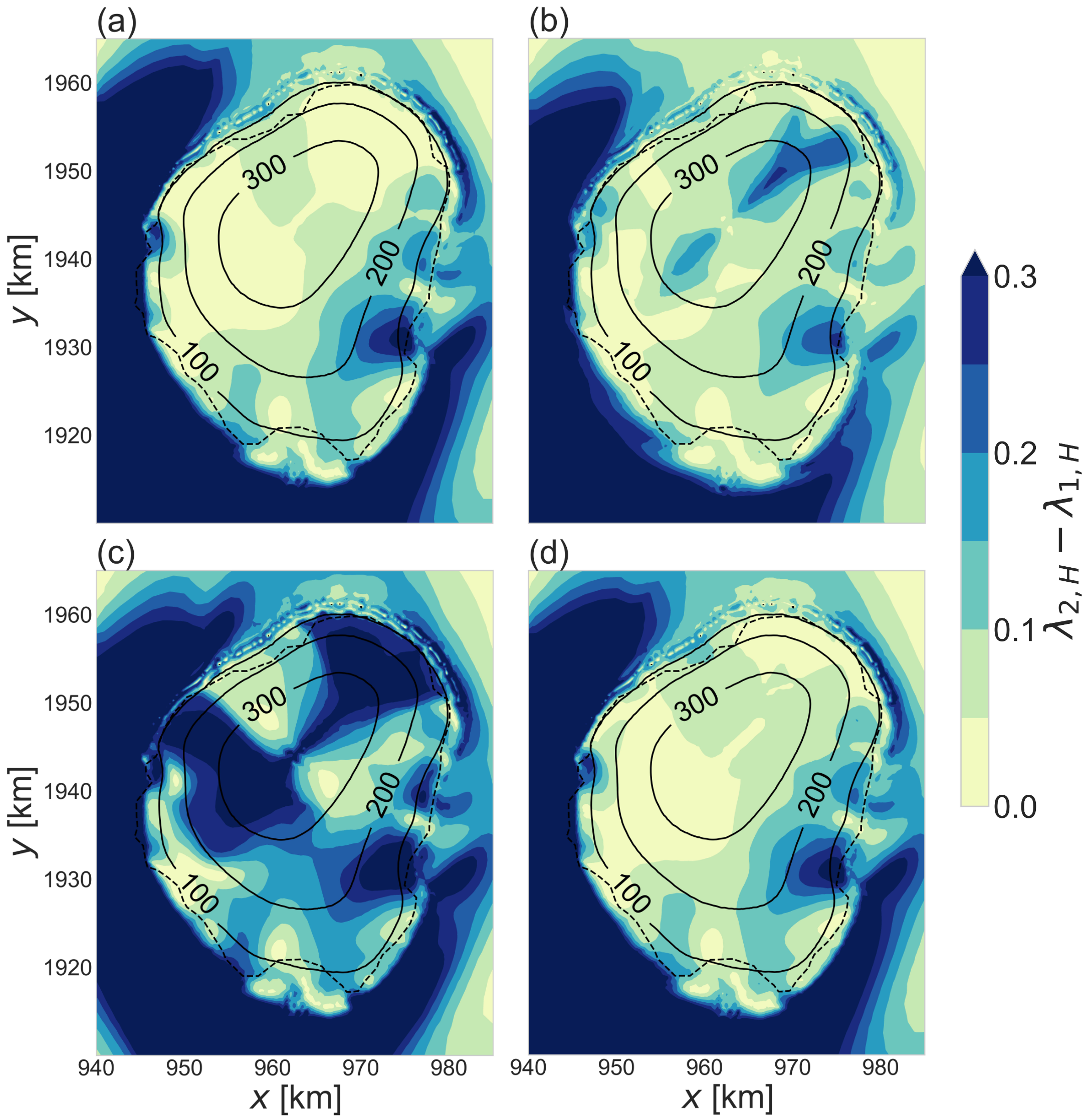}
\caption{The difference between the two eigenvalues of the $2 \times 2$ horizontal crystal orientation tensor for (a) the $\alpha = 0$, $\iota = 1$ simulation, (b) the $\alpha = 0$, $\iota = 0.6$ simulation, (c) the $\alpha = 1$, $\iota = 1$ simulation and (d) the $\alpha = 0.06$, $\iota = 1$ simulation. The solid lines show contours for the depth below the surface in metres and the dashed line is the grounding line.}
\label{a_2H_1H}
\end{figure}

Next, we present results for the difference between the eigenvalues, $\lambda_{2,H} - \lambda_{1,H}$, of the $2 \times 2$ horizontal part of the $3 \times 3$ crystal orientation tensor. This metric corresponds directly with results from radar data where a vertical eigenvector assumption is made (Fig. \ref{a_2H_1H}). The $\alpha=0$, $\iota=1$ (Fig. \ref{a_2H_1H}a) and the $\alpha=0.06$, $\iota=1$ (Fig. \ref{a_2H_1H}d) simulations show similar results both spatially and in terms of the magnitude of differences, with differences staying below $0.1$ at and in the vicinity of the flow divide. The largest differences in the eigenvalues occur on the eastern side of the ice rise, close to the grounding line as well as in small isolated areas elsewhere on the stoss side of the ice rise in the vicinity of the grounding line. Although eigenvalue differences differ substantially in magnitude between the two simulations, the $\alpha=0$, $\iota=0.6$ (Fig. \ref{a_2H_1H}b) and the $\alpha=1$, $\iota=1$ (Fig. \ref{a_2H_1H}c) simulations show similar spatial patterns. In the $\alpha=0$, $\iota=0.6$ simulation (Fig. \ref{a_2H_1H}b), the highest differences between $\lambda_{2,H}$ and $\lambda_{1,H}$ area seen at the tails of the flow divide, extending towards the grounding line north-east of the ice rise and in isolated areas on the east of the ice rise. Elsewhere, differences between $\lambda_{2,H}$ and $\lambda_{1,H}$ remain below $0.1$. The $\alpha=1$, $\iota=1$ simulation (Fig. \ref{a_2H_1H}c) shows large differences between $\lambda_{2,H}$ and $\lambda_{1,H}$ of over $0.3$ at the flow divide and extending to the grounding line, particularly on the lee side of the ice rise. As in all other simulations, differences of over $0.3$ are seen on the eastern side of the ice rise close to the grounding line.

Finally, we investigate the dependence of the anisotropy field on the strain rate field, and in particular the relationship between the largest horizontal anisotropy direction and the largest horizontal strain rate direction. Given that we expect $c$-axes to point perpendicular to the direction of maximum stretching, we calculate the largest horizontal strain rate direction by calculating the eigenvalues and eigenvectors of the upper left $2 \times 2$ tensor of the $3 \times 3$ strain rate tensor, $\dot{\bm{\varepsilon}}$. We denote the eigenvectors of the horizontal strain rate tensor by $\bm{w}_{1,H}$ and $\bm{w}_{2,H}$, where $\bm{w}_{1,H} \leq \bm{w}_{2,H}$. We then calculate the angle between the maximum horizontal strain rate direction, $\bm{w}_{2,H}$, and the maximum horizontal anisotropy direction, $\bm{v}_{2, H}$ (Fig. \ref{a_d_diff}). In the $\alpha=0$, $\iota=1$ simulation (Fig. \ref{a_d_diff}a) and the $\alpha=0.06$, $\iota=1$ simulation (Fig. \ref{a_d_diff}d), the angle between $\bm{w}_{2,H}$ and $\bm{v}_{2, H}$ remains close to $\pi / 2$, meaning that the two vectors are mostly close to perpendicular. Exceptions are in small areas of divergence at the two tails of the flow divide. In the $\alpha=0$, $\iota=0.6$ simulation (Fig. \ref{a_d_diff}b), the angle between $\bm{w}_{2,H}$ and $\bm{v}_{2, H}$ is mainly high everywhere on the ice rise except at the two tails of the flow divide where values are consistently closer to zero. Lastly, the $\alpha=1$, $\iota=1$ simulation (Fig. \ref{a_d_diff}c) shows a large area on the stoss side of the ice rise, away from the flow divide, where $\bm{w}_{2,H}$ and $\bm{v}_{2, H}$ are mostly close to perpendicular. At the flow divide, there is a thin line where the two vectors are perpendicular and in the areas perpendicular to the flow divide, the angle between the two vectors is generally $> \pi / 4$. This simulation shows the largest areas where the angle between $\bm{w}_{2,H}$ and $\bm{v}_{2, H}$ is close to zero from the extremes of the flow divide as far, or almost as far as the grounding line.

\section{Discussion}

We present the results of three-dimensional simulations of the full-tensor anisotropy field of Derwael Ice Rise via a coupling with the velocity and stress fields. The simulated domain includes the surrounding ice shelf, allowing analysis of the anisotropy field across the grounding zone. Based on four previous studies \citep{Martin2009, Seddik2011, Martín2012, Gagliardini2013}, we simulate the anisotropy field varying parameter choices controlling the relative influence of the strain rate and deviatoric stress tensors. By decomposition of the simulated crystal orientation tensor into eigenvalues and eigenvectors, our results provide a necessary step towards the comparison of observed ice fabric with three-dimensional modelled results for a variety of flow regimes.

\subsection{Dependence of the anisotropy field on the velocity and stress fields}

The various flow regimes in the ice rise influence the anisotropy field in different ways, with past flow regimes of a volume of ice also having an effect on the present anisotropy, which is captured due to the semi-Lagrangian implementation of the crystal orientation tensor evolution in the simulations in this study. 

Based on previous choices \citep{Martin2009, Seddik2011, Martín2012, Gagliardini2013}, we adjust the parameters $\alpha$ and $\iota$ in Eq. (\ref{evolution}). The parameter $\alpha$, taking values $\alpha \in [0, 1]$, controls the relative influence of the strain rate and deviatoric stress tensors, with a value of $\alpha = 0$ meaning that the deviatoric stress tensor is ignored and a value of $\alpha=1$ meaning that the strain rate tensor is ignored. The parameter $\iota$, taking values of $\iota \in [0, 1]$, controls the rate of change the crystal orientation tensor, $\bm{a}^{(2)}$, undergoes in response to the combination of the strain rate and deviatoric stress tensor. For example, for a value of $\iota$ close to zero, apart from rotational changes due to the spin tensor, the crystal orientation tensor changes minimally in response to the combination of the strain rate and deviatoric stress tensors. Because the strain rate and deviatoric stress tensor are co-dependent via Glen's flow law (Eq. (\ref{glen})), differentiation between the various combinations of $\alpha$ and $\iota$ is not trivial. Broadly speaking, though this is not always the case, the behaviour of the anisotropy field in the $\alpha=0$, $\iota=0.6$ simulation shows similar results to the $\alpha=1$, $\iota=1$ simulation, despite the relatively large differences in parameter choice. Conversely, the $\alpha=0$, $\iota=1$ and the $\alpha=0.06$, $\iota=1$ simulations show only slightly differing results, likely explained by the similarity in values of $\alpha$ and $\iota$.

\begin{figure}
\center\noindent\includegraphics[width=0.6\textwidth]{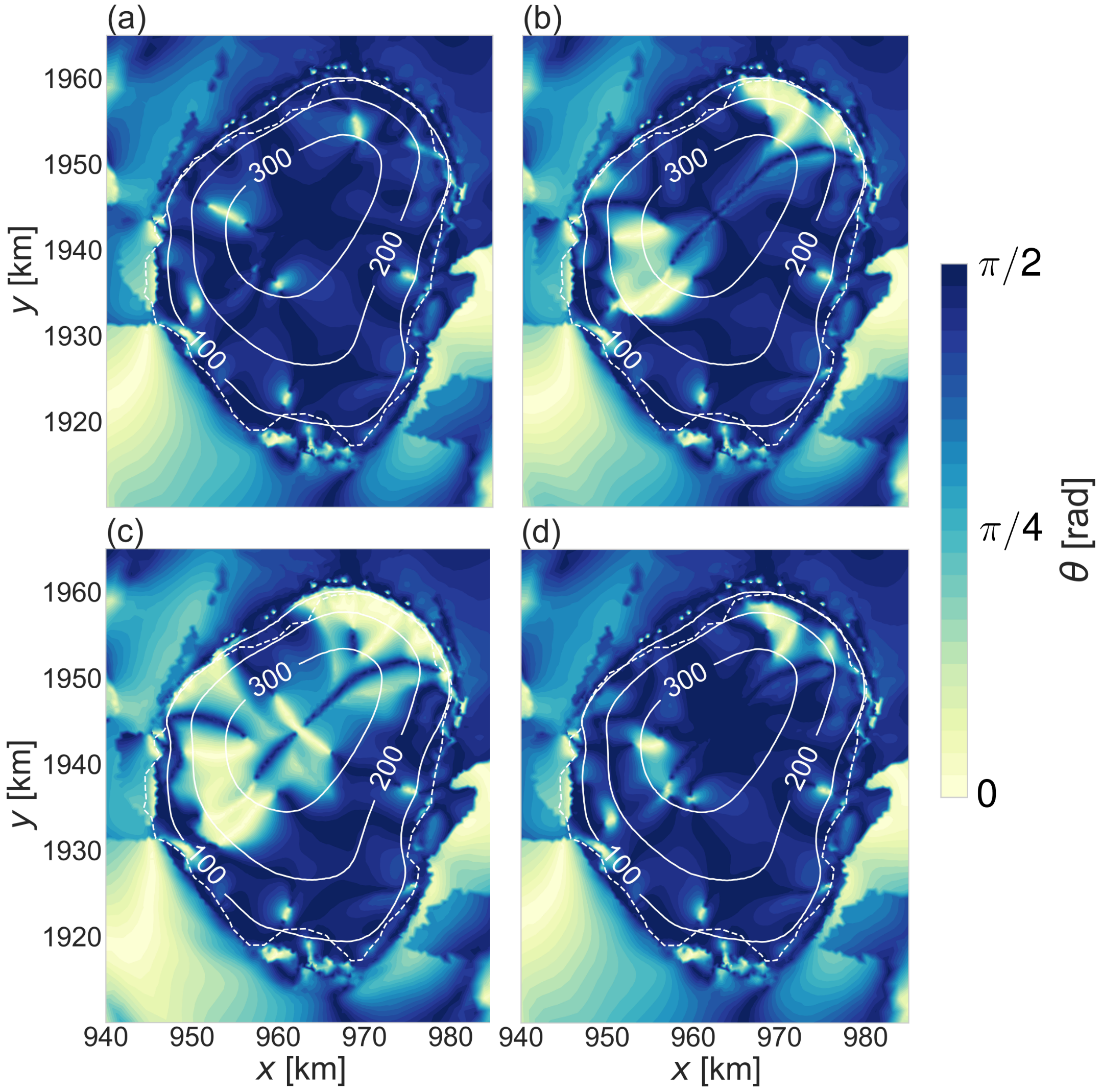}
\caption{The angle between the eigenvectors corresponding to the larger horizontal crystal orientation eigenvalue and the larger horizontal strain rate eigenvalue for (a) the $\alpha = 0$, $\iota = 1$ simulation, (b) the $\alpha = 0$, $\iota = 0.6$ simulation, (c) the $\alpha = 1$, $\iota = 1$ simulation and (d) the $\alpha = 0.06$, $\iota = 1$ simulation.}
\label{a_d_diff}
\end{figure}

While certain anisotropic fabric types can be expected under particular flow and stress regimes, the length of time the ice has undergone deformation in this regime also has an influence on the reflected anisotropy field. In Fig. \ref{lambda_3_2}, it can be seen that at the elevation $z=0$, the ice has developed into a relatively strong single maximum fabric in the $\alpha=0$, $\iota=1$ simulation (Fig. \ref{lambda_3_2}a) and the $\alpha=0.06$, $\iota=1$ simulation (Fig. \ref{lambda_3_2}d) compared with the other two simulations. On the other hand, the $\alpha=1$, $\iota=1$ simulation (Fig. \ref{lambda_3_2}c) does not show a strong single maximum. This indicates that the strain rate term of Eq. (\ref{C}) has a stronger influence on the development of a single maximum under vertical compression at a flow divide than the deviatoric stress tensor term of Eq. (\ref{C}). Interestingly, in the $\alpha=0$, $\iota=1$ simulation (Fig. \ref{lambda_3_2}a) and the $\alpha=0.06$, $\iota=1$ simulation (Fig. \ref{lambda_3_2}d), the areas with a higher single maximum are located at the two tails of the flow divide. Although not explicitly simulated in this study, we expect that the fabric type in the area of high horizontal divergence at the tails of the flow divide would show a similar fabric to that forming at a dome flow divide, defined by a flow divide with horizontal flowlines originating at a point source. In the vicinity of the flow divide, the $\alpha=1$, $\iota=1$ simulation shows a vastly different horizontal anisotropy direction than the other simulations (Fig. \ref{lambda_2_1}c). The eigenvector corresponding to the largest horizontal eigenvalue originate at a point source as opposed to being parallel to the flow divide in the other simulations. Values of $\ln (\lambda_2 / \lambda_1)$ are high along the flow divide and in relatively large areas where there is high horizontal divergence at the tails of the flow divide meaning that the ice has a girdle fabric. Of note is that Fig. \ref{a_d_diff}c shows that in the areas of girdle fabric, the maximum horizontal strain rate direction is not perpendicular to the maximum horizontal anisotropy direction. The reason for the differences between the $\alpha=1$, $\iota=1$ and the other three simulations is due to the largely different $\alpha$ value, which in this case results in ignoring the strain rate tensor completely.

Moving into the flanks of the ice rise, where vertical shear dominates, although the flow of ice in the grounded areas is horizontally divergent, the fabric types vary substantially. We identify a number of factors as having an influence on the ice fabric type: (a) whether flank flow is perpendicular to the flow divide or at the tails of the flow divide, (b) whether the ice is on the stoss or lee side of the ice rise, and (c) how close ice is to the grounding line and what type of flow regime is active at the grounding zone. In flow regimes dominated by vertical shear, as is the case in the areas perpendicular to the flow divide, it is expected that fabric shows a single maximum with a slight offset to the vertical \citep{Llorens2022}. However, the $\alpha=0$, $\iota=1$ (Fig. \ref{lambda_3_2}a) and the $\alpha=0.06$, $\iota=1$ (Fig. \ref{lambda_3_2}d) simulations show the strongest single maxima perpendicular to the flow divide, but Figs. \ref{angle_vert}a and d, show that these simulations have the smallest tilt angle compared to the other simulations.

Across the grounding line, the predicted ln(k) values allow differentiation between single maximum fabric $(\ln (k) > 0)$ and girdle fabrics $(\ln (k) < 0$ , Fig. \ref{lnk}). It must be noted, however, that a large positive or negative value of $\ln (k)$ does not necessarily imply that the ice has a strong single maximum or strong girdle fabric, respectively. Any analysis of the $\ln (k)$ field must be done in conjunction with plots of $\ln (\lambda_2 / \lambda_1)$ (Fig. \ref{lambda_2_1}) and $\ln (\lambda_3 / \lambda_2)$ (Fig. \ref{lambda_3_2}) \citep{Woodcock1977}. In all simulations, ($\ln (\lambda_3 / \lambda_2)$) is the lowest at the grounding zone and in the northern and eastern sides of the ice shelf, meaning that there is no single maximum in these areas. Contrary to this, the metric for a girdle fabric ($\ln (\lambda_2 / \lambda_1)$) shows a small band at the grounding zone with values of $\ln (\lambda_2 / \lambda_1) = 0.5$ to $1$. This weak girdle fabric may be due to the transition from a fabric with a vertical single maximum, to a single maximum aligned with the horizontal plane, via a transition across the grounding zone.

\subsection{Implications for viscosity coupling}

One ultimate goal in modelling the anisotropy field of ice is to be able to fully couple the velocity and stress fields to the anisotropy field and \textit{vice versa} (note, we will refer to these simulations as \textit{fully-coupled} hereafter). In this study, we have explored the influence of the velocity and stress fields on the anisotropy field using three-dimensional simulations, but have not investigated the coupling of the anisotropy field to the velocity field via a direction-dependent viscosity. Here, we discuss areas of Derwael Ice Rise where we expect the fabric type to influence viscosity as well as barriers to fully-coupled, three-dimensional anisotropic simulations.

Generally-speaking, ice crystal $c$-axes rotate in a direction which allows for optimal shearing, with studies showing that if ice is highly anisotropic, deformation in the plane perpendicular to the $c$-axis is significantly enhanced compared with deformation in other directions \citep{Duval1983}. Areas where ice has a single maximum will experience ease of deformation parallel to basal planes and stiffening in response to deformation parallel to the $c$-axes. In the simulations of Derwael Ice Rise, a single maximum fabric is most apparent in the $\alpha=0$, $\iota=1$ and the $\alpha=0.06$, $\iota=1$ simulations (Fig. \ref{lambda_3_2}). If enhancement factors were used \citep{Gillet2005, Ma2010} in a coupling with the ice viscosity, this would allow more ease of vertical shear in these areas which is important in the evolution of Raymond arches at flow divides, as shown in two-dimensional, anisotropic simulations \citep{Martin2009, Martín2012, Drews2015}.

A number of hurdles hinder the three-dimensional simulation of a fully-coupled anisotropy field including computational expense, numerical instabilities, a lack of comparison with observations, parameter uncertainty and challenges associated with parallelisation. In a study by \cite{Gerber2023} in which flow tube simulations the Elmer/Ice model of a fully-coupled aniso\-tropy field were performed, it was noted that numerical instabilities significantly hindered the simulation of areas with high velocities. Furthermore, computational expense and an increased memory load means that a direction-dependent viscosity creates a further hindrance. Despite this, simulations such as those presented in this paper allow ease of comparison with observations and should be included in comparisons between a hierarchy of models with varying complexity.

\subsection{Model representation of anisotropy}

Ice rises are an ideal location for testing models of anisotropy as they contain various flow regimes which are relevant for other locations in the Antarctic Ice Sheet. Ice rises contain many of the same features as entire ice sheets, and results can inform expectations on studies in larger scales \citep{Matsuoka2003}. The shear zones on the western and eastern sides of Derwael Ice Rise are representative of behaviour typical at lateral shear zones in ice shelves or ice streams \citep{Smith2017}. Similarly, flow divides, flank flow and ice shelf flow are typical in many areas of Antarctica. More studies are needed to examine whether an anisotropic flow law is needed in order to make accurate projections of sea level rise. This is especially important as sudden speed up of ice, for example in the initiation of an ice stream, is associated with anisotropic behaviour \citep{Thorsteinsson2003, Lilien2021, Gerber2023}. This necessitates further studies for comparison between various hierarchies of isotropic and anisotropic ice flow models with varying model complexity as well as comparison with observations. In our approach, anisotropy is modelled using a crystal orientation tensor, which can describe many fabric types, but does not capture cone fabric shapes as modelled by \cite{Pettit2007} or multiple single maxima. An example of a representation of crystal orientation which can capture more complex fabrics has been developed by \cite{Rathmann2021}.

In addition to difficulties in choosing a correct mathematical representation of an anisotropic fabric, Eq. (\ref{evolution}) is challenging to solve numerically as it is an advection equation, with no diffusive term. We find that the use of semi-Lagrangian methods reduce dispersion, as they introduce diffusion through interpolation \citep{Advani1987}. However, in areas with strong gradients in the velocity field we nonetheless find that numerical dispersion deteriorates the solution with time. The alternative approach to this numerical problem has been to either explicitly incorporate numerical diffusion in Eq. (\ref{evolution}) as in \cite{Seddik2011}, or to include re-crystallisation terms as in \cite{Gagliardini2013}. We find it to be more effective to constrain the degree of orientation, but our approach can easily be adapted to include a 
re-crystallisation term in Eq. (\ref{evolution}).

\subsection{Framework for comparison with observations}

A crucial component of accurate modelling of the anisotropy field of ice is the comparison with observations, which, until recently have been possible only by comparison with a limited number of ice cores. From ice core data, it is possible to construct a crystal orientation tensor which is directly comparable to simulations such as those presented in this work. Recently, advances in the measurement of anisotropic fabric using radar has allowed the acquisition of a significant amount of data with much greater ease than expensive ice coring projects \citep{Young2021b, Young2021, Ershadi2022}. However, assumptions made in the processing of data mean that the methods currently used are only valid where the dominant $c$-axis direction is exactly vertical \citep{Rathmann2022b}.

Thus far, observational anisotropy data has been used for inferring ice flow history including flow re-organisation \citep{Durand2007, Matsuoka2012, Brisbourne2019, Brisbourne2021} as well as for investigation of the state of the ice fabric in ice conditions such as ice streams \citep{Smith2017, Kufner2023}, in layers of enhanced shear \citep{Pettit2011} and at ice rises \citep{Drews2015}. In terms of making comparisons between observations and anisotropic models, matching of isochrones and velocities has been performed \citep{Drews2015, McCormack2022}, but few studies have compared directly the observed and modelled anisotropy fields \citep{Lilien2023}.

Although we do not explicitly make comparisons between the observed and modelled anisotropy fields of Derwael Ice Rise, we have described steps necessary for such a comparison. Our results show very differing three-dimensional anisotropy fields depending on the chosen influence of the strain rate and deviatoric stress tensors in Eq. (\ref{evolution}) used in previous studies \citep{Martin2009, Seddik2011, Martín2012, Gagliardini2013} and draw attention to the need to constrain equation parameters. We have provided a method to compare the difference between the two horizontal eigenvalues (Fig. \ref{a_2H_1H}), sometimes referred to as the \textit{horizontal anisotropy}, which is directly comparable to radar observations. We recommend radar surveys in the areas of high horizontal flow divergence at the tails of flow divides, as these are areas where differing values of $\alpha$ and $\iota$ show significantly differing anisotropy fields. 

\section{Conclusions}

In this study, we have modelled the three-dimensional anisotropy field of Derwael Ice Rise (DIR) including domain partitioning using a crystal orientation tensor evolution equation describing the spatial distribution of ice crystal $c$-axes in a given ice volume. Simulations are performed varying the relative influence of the strain rate and deviatoric stress tensors in the crystal orientation tensor evolution equation as used in four previous studies \citep{Martin2009, Seddik2011, Martín2012, Gagliardini2013}, with results showing significantly differing ice fabrics across the flow divide and the flanks of DIR. Lastly, we provide a modelling framework for comparison with radar observations, outlining areas where the assumption of one vertical eigenvector may not hold and resulting errors in horizontal eigenvalues.

\section{Acknowledgements} 
C. Henry was supported by the Deutsche Forschungsgemeinschaft (DFG) in the framework of the priority programme 1158 "Antarctic Research with comparative investigations in Arctic ice areas" by a grant SCHA 2139/1-1. R. Drews was partially supported by an Emmy Noether Grant of the Deutsche Forschungsgemeinschaft (DR 822/3-1). This work used resources of the Deutsches Klimarechenzentrum (DKRZ) granted by its Scientific Steering Committee (WLA). We would like to thank Luca Schmidt for comments which improved the manuscript.

\section{Author contribution}

ACJH and CM designed the study, ACJH performed the simulations with input from CM, ACJH performed the analysis with input from CM, ACJH wrote the manuscript which was edited by CM and RD.

\section{Code and data availability}

The code needed to perform the simulations and create the presented in the figures here are available in a publicly-accessible, permanent repository \citep{Henry2024CodeAniso}. The accompanying model input and output data, as well as the processed data are also available in a publicly-accessible, permanent repository \citep{Henry2024DataAniso}. The simulations presented here are dependent on the results of simulations presented in a previous study \citep{Henry2023}, for which code and data are available \citep{Henry2024CodeStrat,Henry2024DataStrat}.

\emergencystretch=1em
\bibliography{main.bib}

\begin{thebibliography}{60}
\providecommand{\natexlab}[1]{#1}
\providecommand{\url}[1]{\texttt{#1}}
\expandafter\ifx\csname urlstyle\endcsname\relax
  \providecommand{\doi}[1]{doi: #1}\else
  \providecommand{\doi}{doi: \begingroup \urlstyle{rm}\Url}\fi

\bibitem[Advani and Tucker(1987)]{Advani1987}
S.~G. Advani and I.~Tucker, Charles~L.
\newblock {The Use of Tensors to Describe and Predict Fiber Orientation in
  Short Fiber Composites}.
\newblock \emph{Journal of Rheology}, 31\penalty0 (8):\penalty0 751--784, 11
  1987.
\newblock \doi{10.1122/1.549945}.

\bibitem[Alley(1988)]{Alley1988}
R.~B. Alley.
\newblock Fabrics in polar ice sheets: development and prediction.
\newblock \emph{Science}, 240\penalty0 (4851):\penalty0 493--495, 1988.

\bibitem[Brisbourne et~al.(2019)Brisbourne, Martín, Smith, Baird, Kendall, and
  Kingslake]{Brisbourne2019}
A.~M. Brisbourne, C.~Martín, A.~M. Smith, A.~F. Baird, J.~M. Kendall, and
  J.~Kingslake.
\newblock Constraining {Recent} {Ice} {Flow} {History} at {Korff} {Ice} {Rise},
  {West} {Antarctica}, {Using} {Radar} and {Seismic} {Measurements} of {Ice}
  {Fabric}.
\newblock \emph{Journal of Geophysical Research: Earth Surface}, 124\penalty0
  (1):\penalty0 175--194, 2019.
\newblock \doi{10.1029/2018JF004776}.

\bibitem[Brisbourne et~al.(2021)Brisbourne, Kendall, Kufner, Hudson, and
  Smith]{Brisbourne2021}
A.~M. Brisbourne, M.~Kendall, S.-K. Kufner, T.~S. Hudson, and A.~M. Smith.
\newblock Downhole distributed acoustic seismic profiling at skytrain ice rise,
  west antarctica.
\newblock \emph{The Cryosphere}, 15\penalty0 (7):\penalty0 3443–3458, 2021.
\newblock \doi{10.5194/tc-15-3443-2021}.

\bibitem[Callens et~al.(2016)Callens, Drews, Witrant, Philippe, and
  Pattyn]{Callens2016}
D.~Callens, R.~Drews, E.~Witrant, M.~Philippe, and F.~Pattyn.
\newblock Temporally stable surface mass balance asymmetry across an ice rise
  derived from radar internal reflection horizons through inverse modeling.
\newblock \emph{Journal of Glaciology}, 62\penalty0 (233):\penalty0 525--534,
  2016.
\newblock \doi{10.1017/jog.2016.41}.

\bibitem[Chung and Kwon(2002)]{Chung2002}
D.~H. Chung and T.~H. Kwon.
\newblock Invariant-based optimal fitting closure approximation for the
  numerical prediction of flow-induced fiber orientation.
\newblock \emph{Journal of Rheology}, 46\penalty0 (1):\penalty0 169–194,
  2002.
\newblock \doi{10.1122/1.1423312}.

\bibitem[Dall(2010)]{Dall2010}
J.~Dall.
\newblock Ice sheet anisotropy measured with polarimetric ice sounding radar.
\newblock In \emph{2010 IEEE International Geoscience and Remote Sensing
  Symposium}, pages 2507--2510, 2010.
\newblock \doi{10.1109/IGARSS.2010.5653528}.

\bibitem[Diez et~al.(2014)Diez, Eisen, Weikusat, Eichler, Hofstede, Bohleber,
  Bohlen, and Polom]{Diez2014}
A.~Diez, O.~Eisen, I.~Weikusat, J.~Eichler, C.~Hofstede, P.~Bohleber,
  T.~Bohlen, and U.~Polom.
\newblock Influence of ice crystal anisotropy on seismic velocity analysis.
\newblock \emph{Annals of Glaciology}, 55\penalty0 (67):\penalty0 97–106,
  2014.
\newblock \doi{10.3189/2014AoG67A002}.

\bibitem[Drews et~al.(2012)Drews, Eisen, Steinhage, Weikusat, Kipfstuhl, and
  Wilhelms]{Drews2012}
R.~Drews, O.~Eisen, D.~Steinhage, I.~Weikusat, S.~Kipfstuhl, and F.~Wilhelms.
\newblock Potential mechanisms for anisotropy in ice-penetrating radar data.
\newblock \emph{Journal of Glaciology}, 58\penalty0 (209):\penalty0 613–624,
  2012.
\newblock \doi{10.3189/2012JoG11J114}.

\bibitem[Drews et~al.(2015)Drews, Matsuoka, Martín, Callens, Bergeot, and
  Pattyn]{Drews2015}
R.~Drews, K.~Matsuoka, C.~Martín, D.~Callens, N.~Bergeot, and F.~Pattyn.
\newblock Evolution of {Derwael} {Ice} {Rise} in {Dronning} {Maud} {Land},
  {Antarctica}, over the last millennia.
\newblock \emph{Journal of Geophysical Research: Earth Surface}, 120\penalty0
  (3):\penalty0 564--579, 2015.
\newblock \doi{10.1002/2014JF003246}.
\newblock URL \url{http://doi.wiley.com/10.1002/2014JF003246}.

\bibitem[Durand et~al.(2007)Durand, Gillet-Chaulet, Svensson, Gagliardini,
  Kipfstuhl, Meyssonnier, Parrenin, Duval, and Dahl-Jensen]{Durand2007}
G.~Durand, F.~Gillet-Chaulet, A.~Svensson, O.~Gagliardini, S.~Kipfstuhl,
  J.~Meyssonnier, F.~Parrenin, P.~Duval, and D.~Dahl-Jensen.
\newblock Change in ice rheology during climate variations--implications for
  ice flow modelling and dating of the epica dome c core.
\newblock \emph{Climate of the Past}, 3\penalty0 (1):\penalty0 155--167, 2007.
\newblock \doi{10.5194/cp-3-155-2007}.

\bibitem[Durand et~al.(2009)Durand, Gagliardini, de~Fleurian, Zwinger, and
  Le~Meur]{Durand2009}
G.~Durand, O.~Gagliardini, B.~de~Fleurian, T.~Zwinger, and E.~Le~Meur.
\newblock Marine ice sheet dynamics: {Hysteresis} and neutral equilibrium.
\newblock \emph{Journal of Geophysical Research}, 114\penalty0 (F3):\penalty0
  F03009, Aug. 2009.
\newblock \doi{10.1029/2008JF001170}.

\bibitem[Duval et~al.(1983)Duval, Ashby, and Anderman]{Duval1983}
P.~Duval, M.~Ashby, and I.~Anderman.
\newblock Rate-controlling processes in the creep of polycrystalline ice.
\newblock \emph{The Journal of Physical Chemistry}, 87\penalty0 (21):\penalty0
  4066--4074, 1983.
\newblock \doi{10.1021/j100244a014}.

\bibitem[Ershadi et~al.(2022)Ershadi, Drews, Martín, Eisen, Ritz, Corr,
  Christmann, Zeising, Humbert, and Mulvaney]{Ershadi2022}
M.~R. Ershadi, R.~Drews, C.~Martín, O.~Eisen, C.~Ritz, H.~Corr, J.~Christmann,
  O.~Zeising, A.~Humbert, and R.~Mulvaney.
\newblock Polarimetric radar reveals the spatial distribution of ice fabric at
  domes and divides in east antarctica.
\newblock \emph{The Cryosphere}, 16\penalty0 (5):\penalty0 1719–1739, 2022.
\newblock \doi{10.5194/tc-16-1719-2022}.

\bibitem[Favier and Pattyn(2015)]{Favier2015}
L.~Favier and F.~Pattyn.
\newblock Antarctic ice rise formation, evolution, and stability.
\newblock \emph{Geophysical Research Letters}, 42\penalty0 (11):\penalty0
  4456--4463, June 2015.
\newblock ISSN 0094-8276, 1944-8007.
\newblock \doi{10.1002/2015GL064195}.

\bibitem[Fujita et~al.(2006)Fujita, Maeno, and Matsuoka]{Fujita2006}
S.~Fujita, H.~Maeno, and K.~Matsuoka.
\newblock Radio-wave depolarization and scattering within ice sheets: a
  matrix-based model to link radar and ice-core measurements and its
  application.
\newblock \emph{Journal of Glaciology}, 52\penalty0 (178):\penalty0 407–424,
  2006.
\newblock \doi{10.3189/172756506781828548}.

\bibitem[Gagliardini et~al.(2009)Gagliardini, Gillet-Chaulet, and
  Montagnat]{Gagliardini2009}
O.~Gagliardini, F.~Gillet-Chaulet, and M.~Montagnat.
\newblock A review of anisotropic polar ice models: from crystal to ice-sheet
  flow models.
\newblock \emph{Physics of Ice Core Records II}, 68, 2009.

\bibitem[Gagliardini et~al.(2013)Gagliardini, Zwinger, Gillet-Chaulet, Durand,
  Favier, de~Fleurian, Greve, Malinen, Martín, Råback, Ruokolainen,
  Sacchettini, Schäfer, Seddik, and Thies]{Gagliardini2013}
O.~Gagliardini, T.~Zwinger, F.~Gillet-Chaulet, G.~Durand, L.~Favier,
  B.~de~Fleurian, R.~Greve, M.~Malinen, C.~Martín, P.~Råback, J.~Ruokolainen,
  M.~Sacchettini, M.~Schäfer, H.~Seddik, and J.~Thies.
\newblock Capabilities and performance of {Elmer}/{Ice}, a new-generation ice
  sheet model.
\newblock \emph{Geoscientific Model Development}, 6\penalty0 (4):\penalty0
  1299--1318, 2013.
\newblock \doi{10.5194/gmd-6-1299-2013}.

\bibitem[Gerber et~al.(2023)Gerber, Lilien, Rathmann, Franke, Young,
  Valero-Delgado, Ershadi, Drews, Zeising, Humbert, et~al.]{Gerber2023}
T.~A. Gerber, D.~A. Lilien, N.~M. Rathmann, S.~Franke, T.~J. Young,
  F.~Valero-Delgado, M.~R. Ershadi, R.~Drews, O.~Zeising, A.~Humbert, et~al.
\newblock Crystal orientation fabric anisotropy causes directional hardening of
  the northeast greenland ice stream.
\newblock \emph{Nature Communications}, 14\penalty0 (1):\penalty0 2653, 2023.
\newblock \doi{10.1038/s41467-023-38139-8}.

\bibitem[Gillet-Chaulet et~al.(2005)Gillet-Chaulet, Gagliardini, Meyssonnier,
  Montagnat, and Castelnau]{Gillet2005}
F.~Gillet-Chaulet, O.~Gagliardini, J.~Meyssonnier, M.~Montagnat, and
  O.~Castelnau.
\newblock A user-friendly anisotropic flow law for ice-sheet modeling.
\newblock \emph{Journal of glaciology}, 51\penalty0 (172):\penalty0 3--14,
  2005.
\newblock \doi{10.3189/172756505781829584}.

\bibitem[Gillet-Chaulet et~al.(2006)Gillet-Chaulet, Gagliardini, Meyssonnier,
  Zwinger, and Ruokolainen]{Gillet2006}
F.~Gillet-Chaulet, O.~Gagliardini, J.~Meyssonnier, T.~Zwinger, and
  J.~Ruokolainen.
\newblock Flow-induced anisotropy in polar ice and related ice-sheet flow
  modelling.
\newblock \emph{Journal of non-newtonian fluid mechanics}, 134\penalty0
  (1-3):\penalty0 33--43, 2006.
\newblock \doi{10.1016/j.jnnfm.2005.11.005}.

\bibitem[Gödert(2003)]{godert2003}
G.~Gödert.
\newblock A mesoscopic approach for modelling texture evolution of polar ice
  including recrystallization phenomena.
\newblock \emph{Annals of Glaciology}, 37:\penalty0 23–28, 2003.
\newblock \doi{10.3189/172756403781815375}.

\bibitem[Henry(2024{\natexlab{a}})]{Henry2024CodeAniso}
A.~C.~J. Henry.
\newblock Simulation and post-processing code presented in "modelling the
  three-dimensional, diagnostic anisotropy field of an ice rise" ({Version
  v1}).
\newblock Zenodo. \url{https://doi.org/10.5281/zenodo.13143326},
  2024{\natexlab{a}}.
\newblock Accessed: 2024-07-31.

\bibitem[Henry(2024{\natexlab{b}})]{Henry2024CodeStrat}
A.~C.~J. Henry.
\newblock Code for the publication ``{Predicting} the three-dimensional
  stratigraphy of an ice rise" ({Version v1}).
\newblock Zenodo. \url{https://doi.org/10.5281/zenodo.12180129},
  2024{\natexlab{b}}.
\newblock Accessed: 2024-06-20.

\bibitem[Henry(2024{\natexlab{c}})]{Henry2024DataAniso}
A.~C.~J. Henry.
\newblock Model input data, simulation output data and processed data presented
  in "modelling the three-dimensional, diagnostic anisotropy field of an ice
  rise" ({Version v1}).
\newblock Zenodo. \url{https://doi.org/10.5281/zenodo.13143591},
  2024{\natexlab{c}}.
\newblock Accessed: 2024-07-31.

\bibitem[Henry(2024{\natexlab{d}})]{Henry2024DataStrat}
A.~C.~J. Henry.
\newblock Data for the publication ``predicting the three-dimensional
  stratigraphy of an ice rise" ({Version v1}).
\newblock Zenodo. \url{https://doi.org/10.5281/zenodo.12183293},
  2024{\natexlab{d}}.
\newblock Accessed: 2024-06-20.

\bibitem[Henry et~al.(2022)Henry, Drews, Schannwell, and
  Vi\v{s}njevi\'c]{Henry2022}
A.~C.~J. Henry, R.~Drews, C.~Schannwell, and V.~Vi\v{s}njevi\'c.
\newblock Hysteretic evolution of ice rises and ice rumples in response to
  variations in sea level.
\newblock \emph{The Cryosphere}, 16\penalty0 (9):\penalty0 3889--3905, 2022.
\newblock \doi{10.5194/tc-16-3889-2022}.

\bibitem[Henry et~al.(2023)Henry, Schannwell, Vi\v{s}njevi\'c, Millstein, Bons,
  Eisen, and R.]{Henry2023}
A.~C.~J. Henry, C.~Schannwell, V.~Vi\v{s}njevi\'c, J.~Millstein, P.~Bons,
  O.~Eisen, and D.~R.
\newblock Predicting the three-dimensional age-depth field of an ice rise.
\newblock 2023.
\newblock \doi{10.22541/essoar.169230234.44865946/v1}.

\bibitem[Jezek et~al.(2013)Jezek, Curlander, Carsey, Wales, and
  Barry.]{Jezek2013}
K.~C. Jezek, J.~C. Curlander, F.~Carsey, C.~Wales, and R.~G. Barry.
\newblock Ramp amm-1 sar image mosaic of antarctica, version 2, 2013.

\bibitem[Jordan et~al.(2022)Jordan, Martín, Brisbourne, Schroeder, and
  Smith]{Jordan22}
T.~M. Jordan, C.~Martín, A.~M. Brisbourne, D.~M. Schroeder, and A.~M. Smith.
\newblock Radar characterization of ice crystal orientation fabric and
  anisotropic viscosity within an antarctic ice stream.
\newblock \emph{Journal of Geophysical Research: Earth Surface}, 127\penalty0
  (6):\penalty0 e2022JF006673, 2022.
\newblock \doi{https://doi.org/10.1029/2022JF006673}.

\bibitem[Koch et~al.(2023)Koch, Drews, Franke, Jansen, Oraschewski, Muhle,
  Višnjević, Matsuoka, Pattyn, and Eisen]{Koch2023}
I.~Koch, R.~Drews, S.~Franke, D.~Jansen, F.~M. Oraschewski, L.~S. Muhle,
  V.~Višnjević, K.~Matsuoka, F.~Pattyn, and O.~Eisen.
\newblock Radar internal reflection horizons from multisystem data reflect ice
  dynamic and surface accumulation history along the princess ragnhild coast,
  dronning maud land, east antarctica.
\newblock \emph{Journal of Glaciology}, page 1–19, 2023.
\newblock \doi{10.1017/jog.2023.93}.

\bibitem[Kufner et~al.(2023)Kufner, Wookey, Brisbourne, Martín, Hudson,
  Kendall, and Smith]{Kufner2023}
S.~Kufner, J.~Wookey, A.~M. Brisbourne, C.~Martín, T.~S. Hudson, J.~M.
  Kendall, and A.~M. Smith.
\newblock Strongly depth‐dependent ice fabric in a fast‐flowing antarctic
  ice stream revealed with icequake observations.
\newblock \emph{Journal of Geophysical Research: Earth Surface}, 128\penalty0
  (3):\penalty0 e2022JF006853, 2023.
\newblock \doi{10.1029/2022JF006853}.

\bibitem[Lilien et~al.(2021)Lilien, Rathmann, Hvidberg, and
  Dahl‐Jensen]{Lilien2021}
D.~A. Lilien, N.~M. Rathmann, C.~S. Hvidberg, and D.~Dahl‐Jensen.
\newblock Modeling ice‐crystal fabric as a proxy for ice‐stream stability.
\newblock \emph{Journal of Geophysical Research: Earth Surface}, 126\penalty0
  (9), 2021.
\newblock \doi{10.1029/2021JF006306}.

\bibitem[Lilien et~al.(2023)Lilien, Rathmann, Hvidberg, Grinsted, Ershadi,
  Drews, and Dahl-Jensen]{Lilien2023}
D.~A. Lilien, N.~M. Rathmann, C.~S. Hvidberg, A.~Grinsted, M.~R. Ershadi,
  R.~Drews, and D.~Dahl-Jensen.
\newblock Simulating higher-order fabric structure in a coupled, anisotropic
  ice-flow model: application to dome c.
\newblock \emph{Journal of Glaciology}, page 1–20, 2023.
\newblock \doi{10.1017/jog.2023.78}.

\bibitem[Llorens et~al.(2022)Llorens, Griera, Bons, Weikusat, Prior,
  Gomez-Rivas, De~Riese, Jimenez-Munt, García-Castellanos, and
  Lebensohn]{Llorens2022}
M.-G. Llorens, A.~Griera, P.~D. Bons, I.~Weikusat, D.~J. Prior, E.~Gomez-Rivas,
  T.~De~Riese, I.~Jimenez-Munt, D.~García-Castellanos, and R.~A. Lebensohn.
\newblock Can changes in deformation regimes be inferred from crystallographic
  preferred orientations in polar ice?
\newblock \emph{The Cryosphere}, 16\penalty0 (5):\penalty0 2009–2024, 2022.
\newblock \doi{10.5194/tc-16-2009-2022}.

\bibitem[Ma et~al.(2010)Ma, Gagliardini, Ritz, Gillet-Chaulet, Durand, and
  Montagnat]{Ma2010}
Y.~Ma, O.~Gagliardini, C.~Ritz, F.~Gillet-Chaulet, G.~Durand, and M.~Montagnat.
\newblock Enhancement factors for grounded ice and ice shelves inferred from an
  anisotropic ice-flow model.
\newblock \emph{Journal of Glaciology}, 56\penalty0 (199):\penalty0 805–812,
  2010.
\newblock \doi{10.3189/002214310794457209}.

\bibitem[Martín and Gudmundsson(2012)]{Martín2012}
C.~Martín and G.~H. Gudmundsson.
\newblock Effects of nonlinear rheology, temperature and anisotropy on the
  relationship between age and depth at ice divides.
\newblock \emph{The Cryosphere}, 6\penalty0 (5):\penalty0 1221–1229, 2012.
\newblock \doi{10.5194/tc-6-1221-2012}.

\bibitem[Martín et~al.(2009)Martín, Hindmarsh, and Navarro]{Martin2009}
C.~Martín, R.~C.~A. Hindmarsh, and F.~J. Navarro.
\newblock On the effects of divide migration, along‐ridge flow, and basal
  sliding on isochrones near an ice divide.
\newblock \emph{Journal of Geophysical Research}, 114\penalty0 (F2):\penalty0
  F02006, Apr. 2009.
\newblock ISSN 0148-0227.
\newblock \doi{10.1029/2008JF001025}.
\newblock URL \url{http://doi.wiley.com/10.1029/2008JF001025}.

\bibitem[Matsuoka et~al.(2003)Matsuoka, Furukawa, Fujita, Maeno, Uratsuka,
  Naruse, and Watanabe]{Matsuoka2003}
K.~Matsuoka, T.~Furukawa, S.~Fujita, H.~Maeno, S.~Uratsuka, R.~Naruse, and
  O.~Watanabe.
\newblock Crystal orientation fabrics within the antarctic ice sheet revealed
  by a multipolarization plane and dual‐frequency radar survey.
\newblock \emph{Journal of Geophysical Research: Solid Earth}, 108\penalty0
  (B10):\penalty0 2003JB002425, 2003.
\newblock \doi{10.1029/2003JB002425}.

\bibitem[Matsuoka et~al.(2009)Matsuoka, Wilen, Hurley, and
  Raymond]{Matsuoka2009}
K.~Matsuoka, L.~Wilen, S.~Hurley, and C.~Raymond.
\newblock Effects of birefringence within ice sheets on obliquely propagating
  radio waves.
\newblock \emph{IEEE Transactions on Geoscience and Remote Sensing},
  47\penalty0 (5):\penalty0 1429–1443, 2009.
\newblock \doi{10.1109/TGRS.2008.2005201}.

\bibitem[Matsuoka et~al.(2012)Matsuoka, Power, Fujita, and
  Raymond]{Matsuoka2012}
K.~Matsuoka, D.~Power, S.~Fujita, and C.~F. Raymond.
\newblock Rapid development of anisotropic ice-crystal-alignment fabrics
  inferred from englacial radar polarimetry, central west antarctica.
\newblock \emph{Journal of Geophysical Research: Earth Surface}, 117\penalty0
  (F3), 2012.
\newblock \doi{https://doi.org/10.1029/2012JF002440}.

\bibitem[Matsuoka et~al.(2015)Matsuoka, Hindmarsh, Moholdt, Bentley, Pritchard,
  Brown, Conway, Drews, Durand, Goldberg, Hattermann, Kingslake, Lenaerts,
  Martín, Mulvaney, Nicholls, Pattyn, Ross, Scambos, and
  Whitehouse]{Matsuoka2015}
K.~Matsuoka, R.~C. Hindmarsh, G.~Moholdt, M.~J. Bentley, H.~D. Pritchard,
  J.~Brown, H.~Conway, R.~Drews, G.~Durand, D.~Goldberg, T.~Hattermann,
  J.~Kingslake, J.~T. Lenaerts, C.~Martín, R.~Mulvaney, K.~W. Nicholls,
  F.~Pattyn, N.~Ross, T.~Scambos, and P.~L. Whitehouse.
\newblock Antarctic ice rises and rumples: {Their} properties and significance
  for ice-sheet dynamics and evolution.
\newblock \emph{Earth-Science Reviews}, 150:\penalty0 724--745, 2015.
\newblock \doi{10.1016/j.earscirev.2015.09.004}.

\bibitem[McCormack et~al.(2022)McCormack, Warner, Seroussi, Dow, Roberts, and
  Treverrow]{McCormack2022}
F.~S. McCormack, R.~C. Warner, H.~Seroussi, C.~F. Dow, J.~L. Roberts, and
  A.~Treverrow.
\newblock Modeling the deformation regime of thwaites glacier, west antarctica,
  using a simple flow relation for ice anisotropy (estar).
\newblock \emph{Journal of Geophysical Research: Earth Surface}, 127\penalty0
  (3):\penalty0 e2021JF006332, 2022.
\newblock \doi{10.1029/2021JF006332}.

\bibitem[Montagnat et~al.(2014)Montagnat, Azuma, Dahl-Jensen, Eichler, Fujita,
  Gillet-Chaulet, Kipfstuhl, Samyn, Svensson, and Weikusat]{Montagnat2014}
M.~Montagnat, N.~Azuma, D.~Dahl-Jensen, J.~Eichler, S.~Fujita,
  F.~Gillet-Chaulet, S.~Kipfstuhl, D.~Samyn, A.~Svensson, and I.~Weikusat.
\newblock Fabric along the neem ice core, greenland, and its comparison with
  grip and ngrip ice cores.
\newblock \emph{The Cryosphere}, 8\penalty0 (4):\penalty0 1129--1138, 2014.
\newblock \doi{10.5194/tc-8-1129-2014}.

\bibitem[Pettit et~al.(2007)Pettit, Thorsteinsson, Jacobson, and
  Waddington]{Pettit2007}
E.~C. Pettit, T.~Thorsteinsson, H.~P. Jacobson, and E.~D. Waddington.
\newblock The role of crystal fabric in flow near an ice divide.
\newblock \emph{Journal of Glaciology}, 53\penalty0 (181):\penalty0 277–288,
  2007.
\newblock \doi{10.3189/172756507782202766}.

\bibitem[Pettit et~al.(2011)Pettit, Waddington, Harrison, Thorsteinsson,
  Elsberg, Morack, and Zumberge]{Pettit2011}
E.~C. Pettit, E.~D. Waddington, W.~D. Harrison, T.~Thorsteinsson, D.~Elsberg,
  J.~Morack, and M.~A. Zumberge.
\newblock The crossover stress, anisotropy and the ice flow law at siple dome,
  west antarctica.
\newblock \emph{Journal of Glaciology}, 57\penalty0 (201):\penalty0 39–52,
  2011.
\newblock \doi{10.3189/002214311795306619}.

\bibitem[Rathmann et~al.(2021)Rathmann, Hvidberg, Grinsted, Lilien, and
  Dahl-Jensen]{Rathmann2021}
N.~M. Rathmann, C.~S. Hvidberg, A.~Grinsted, D.~A. Lilien, and D.~Dahl-Jensen.
\newblock Effect of an orientation-dependent non-linear grain fluidity on bulk
  directional enhancement factors.
\newblock \emph{Journal of Glaciology}, 67\penalty0 (263):\penalty0 569–575,
  2021.
\newblock \doi{10.1017/jog.2020.117}.

\bibitem[Rathmann et~al.(2022)Rathmann, Lilien, Grinsted, Gerber, Young, and
  Dahl‐Jensen]{Rathmann2022b}
N.~M. Rathmann, D.~A. Lilien, A.~Grinsted, T.~A. Gerber, T.~J. Young, and
  D.~Dahl‐Jensen.
\newblock On the limitations of using polarimetric radar sounding to infer the
  crystal orientation fabric of ice masses.
\newblock \emph{Geophysical Research Letters}, 49\penalty0 (1):\penalty0
  e2021GL096244, 2022.
\newblock \doi{10.1029/2021GL096244}.

\bibitem[Richards et~al.(2022)Richards, Pegler, and Piazolo]{Richards2022}
D.~H. Richards, S.~S. Pegler, and S.~Piazolo.
\newblock Ice fabrics in two-dimensional flows: beyond pure and simple shear.
\newblock \emph{The Cryosphere}, 16\penalty0 (10):\penalty0 4571–4592, 2022.
\newblock \doi{10.5194/tc-16-4571-2022}.

\bibitem[Richards et~al.(2023)Richards, Pegler, Piazolo, Stoll, and
  Weikusat]{Richards2023}
D.~H. Richards, S.~S. Pegler, S.~Piazolo, N.~Stoll, and I.~Weikusat.
\newblock Bridging the gap between experimental and natural fabrics: Modeling
  ice stream fabric evolution and its comparison with ice-core data.
\newblock \emph{Journal of Geophysical Research: Solid Earth}, 128\penalty0
  (11):\penalty0 e2023JB027245, 2023.
\newblock \doi{https://doi.org/10.1029/2023JB027245}.

\bibitem[Rignot and Scheuchl.(2017)]{Rignot2017}
J.~M. Rignot, E. and B.~Scheuchl.
\newblock Measures insar-based antarctica ice velocity map, version 2, 2017.
\newblock URL \url{https://nsidc.org/data/NSIDC-0484/versions/2}.

\bibitem[Schannwell et~al.(2019)Schannwell, Drews, Ehlers, Eisen, Mayer, and
  Gillet-Chaulet]{Schannwell2019}
C.~Schannwell, R.~Drews, T.~A. Ehlers, O.~Eisen, C.~Mayer, and
  F.~Gillet-Chaulet.
\newblock Kinematic response of ice-rise divides to changes in ocean and
  atmosphere forcing.
\newblock \emph{The Cryosphere}, 13\penalty0 (10):\penalty0 2673--2691, 2019.
\newblock \doi{10.5194/tc-13-2673-2019}.

\bibitem[Schannwell et~al.(2020)Schannwell, Drews, Ehlers, Eisen, Mayer,
  Malinen, Smith, and Eisermann]{Schannwell2020}
C.~Schannwell, R.~Drews, T.~A. Ehlers, O.~Eisen, C.~Mayer, M.~Malinen, E.~C.
  Smith, and H.~Eisermann.
\newblock Quantifying the effect of ocean bed properties on ice sheet geometry
  over 40 000 years with a full-{Stokes} model.
\newblock \emph{The Cryosphere}, 14\penalty0 (11):\penalty0 3917--3934, 2020.
\newblock \doi{10.5194/tc-14-3917-2020}.

\bibitem[Seddik et~al.(2011)Seddik, Greve, Zwinger, and Placidi]{Seddik2011}
H.~Seddik, R.~Greve, T.~Zwinger, and L.~Placidi.
\newblock A full stokes ice flow model for the vicinity of dome fuji,
  antarctica, with induced anisotropy and fabric evolution.
\newblock \emph{The Cryosphere}, 5\penalty0 (2):\penalty0 495--508, 2011.
\newblock \doi{10.5194/tc-5-495-2011}.

\bibitem[Smith et~al.(2017)Smith, Baird, Kendall, Martín, White, Brisbourne,
  and Smith]{Smith2017}
E.~C. Smith, A.~F. Baird, J.~M. Kendall, C.~Martín, R.~S. White, A.~M.
  Brisbourne, and A.~M. Smith.
\newblock Ice fabric in an antarctic ice stream interpreted from seismic
  anisotropy.
\newblock \emph{Geophysical Research Letters}, 44\penalty0 (8):\penalty0
  3710–3718, 2017.
\newblock \doi{10.1002/2016GL072093}.

\bibitem[Thorsteinsson et~al.(2003)Thorsteinsson, Waddington, and
  Fletcher]{Thorsteinsson2003}
T.~Thorsteinsson, E.~D. Waddington, and R.~C. Fletcher.
\newblock Spatial and temporal scales of anisotropic effects in ice-sheet flow.
\newblock \emph{Annals of Glaciology}, 37:\penalty0 40–48, 2003.
\newblock \doi{10.3189/172756403781815429}.

\bibitem[Weikusat et~al.(2017)Weikusat, Jansen, Binder, Eichler, Faria,
  Wilhelms, Kipfstuhl, Sheldon, Miller, Dahl-Jensen, and Kleiner]{Weikusat2017}
I.~Weikusat, D.~Jansen, T.~Binder, J.~Eichler, S.~H. Faria, F.~Wilhelms,
  S.~Kipfstuhl, S.~Sheldon, H.~Miller, D.~Dahl-Jensen, and T.~Kleiner.
\newblock Physical analysis of an antarctic ice core—towards an integration
  of micro- and macrodynamics of polar ice.
\newblock \emph{Philosophical Transactions of the Royal Society A:
  Mathematical, Physical and Engineering Sciences}, 375\penalty0
  (2086):\penalty0 20150347, 2017.
\newblock \doi{10.1098/rsta.2015.0347}.

\bibitem[Woodcock(1977)]{Woodcock1977}
N.~H. Woodcock.
\newblock Specification of fabric shapes using an eigenvalue method.
\newblock \emph{Geological Society of America Bulletin}, 88\penalty0
  (9):\penalty0 1231, 1977.
\newblock \doi{10.1130/0016-7606(1977)88<1231:SOFSUA>2.0.CO;2}.

\bibitem[Young et~al.(2021{\natexlab{a}})Young, Mart{\'\i}n, Christoffersen,
  Schroeder, Tulaczyk, and Dawson]{Young2021b}
T.~J. Young, C.~Mart{\'\i}n, P.~Christoffersen, D.~M. Schroeder, S.~M.
  Tulaczyk, and E.~J. Dawson.
\newblock Rapid and accurate polarimetric radar measurements of ice crystal
  fabric orientation at the western antarctic ice sheet (wais) divide ice core
  site.
\newblock \emph{The Cryosphere}, 15\penalty0 (8):\penalty0 4117--4133,
  2021{\natexlab{a}}.

\bibitem[Young et~al.(2021{\natexlab{b}})Young, Schroeder, Jordan,
  Christoffersen, Tulaczyk, Culberg, and Bienert]{Young2021}
T.~J. Young, D.~M. Schroeder, T.~M. Jordan, P.~Christoffersen, S.~M. Tulaczyk,
  R.~Culberg, and N.~L. Bienert.
\newblock Inferring ice fabric from birefringence loss in airborne radargrams:
  Application to the eastern shear margin of thwaites glacier, west antarctica.
\newblock \emph{Journal of Geophysical Research: Earth Surface}, 126\penalty0
  (5):\penalty0 e2020JF006023, 2021{\natexlab{b}}.
\newblock \doi{10.1029/2020JF006023}.

\end{thebibliography}

\end{document}